\documentstyle{mn}

\input{epsf.sty}

\def\plottwo#1#2{\centering \leavevmode
\epsfxsize=.45\columnwidth \epsfbox{#1} \hfil
\epsfxsize=.45\columnwidth \epsfbox{#2}}

\def\plotthree#1#2#3{\centering \leavevmode
\epsfxsize=.32\columnwidth \epsfbox{#1} \hfil
\epsfxsize=.32\columnwidth \epsfbox{#2} \hfil
\epsfxsize=.32\columnwidth \epsfbox{#3}}

\def\plotxsize#1#2{\centering \leavevmode \epsfxsize=#2 \epsfbox{#1}}

\def\plotysize#1#2{\centering \leavevmode \epsfysize=#2 \epsfbox{#1}}

\def\plotps#1{\centering \leavevmode \epsfbox{#1}}

\newcommand{\ff}         {_{\mbox{\tiny ff}}}
\newcommand{\smass}      {\mbox{M}_{\odot}}
\newcommand{\pcc}        {cm$^{-3}$}
\newcommand{\kms}        {km s$^{-1}$}
\newcommand{\iso}        {_{\mbox{\tiny iso}}}
\newcommand{\logn}       {\log_{10} (N/\mbox{cm}^{-2})}
\newcommand{\bsw}        {b_{\mbox{\tiny switch}}}

\begin{document}

\title[Dynamically triggered star formation in giant molecular clouds]
      {Dynamically triggered star formation in giant molecular clouds}

\author[A.S.~Bhattal et al.]
       {A.S.~Bhattal, N.~Francis, S.J.~Watkins, and A.P.~Whitworth\\
       Department of Physics and Astronomy, University of Wales, Cardiff CF2 3YB, Wales, UK.}

\maketitle

\begin{abstract}
A Lagrangian, particle-based numerical method (tree code gravity plus smoothed particle hydrodynamics) was used to simulate clump-clump collisions occurring within GMCs. The collisions examined were between 75$\smass$ clumps at a relative Mach number of ${\cal M}=3$. The clumps were modelled using isothermal spheres which were individually in stable equilibrium.

The collisions formed shock-compressed layers, out of which condensed approximately co-planar protostellar discs of 7-60$\smass$ mass and 500-1000AU radius. Binary and multiple systems were the usual final state. Lower mass objects were also produced, but commonly underwent disruption or merger. Such objects occasionally survived by being ejected via a three-body slingshot event resulting from an encounter with a binary system.

The impact parameter $b$ denotes how offset the clumps are from one another, with low values corresponding to near-head on collisions, and high values corresponding to grazing collisions. Varying $b$ altered the processes by which the protostellar systems formed. At low $b$ a single central disc formed initially, and was then spun-up by an accretion flow, causing it to produce secondaries via rotational instabilities. At mid $b$ the shocked layer which formed initially broke up into fragments, and discs were then formed via fragment merger. At large $b$ single objects formed within the compressed leading edge of each clump. These became unbound from each other as $b$ was increased further.

The effect of changing numerical factors was examined by : (i) colliding clumps which had been re-oriented before the collision (thus altering the initial particle noise), and (ii) by quadrupling the number of particles in each clump (thus increasing the resolution of the simulation). Both changes were found to affect the small-scale details of a collision, but leave the large scale morphology largely unaltered.

It was concluded that clump-clump collisions provide a natural mechanism by which multiple protostellar systems may form.
\end{abstract}


\section{Introduction.}
\label{sec:intro}

This paper proceeds from the assumption that most star formation is dynamically triggered and produces clusters, rather than occuring spontaneously in a more widely distributed mode. We explore the mechanics of star formation triggered by clump-clump collisions within molecular clouds. Such collisions can occur either within a single molecular cloud, as a result of the random velocities of the constituent clumps, or when two molecular clouds collide and the collision is broken up into an ensemble of approximately simultaneous clump-clump collisions.

If the sub-structure within molecular clouds is hierarchical, as has been inferred \cite{sca}, the most common collisions will occur between clumps at similar levels of the hierarchy (and hence between clumps of similar mass).  According to perturbation theory \cite{whi-lay}, the characteristic masses of fragments condensing out of a shock-compressed layer are determined mainly by the post-shock temperature $T_{s}$. For typical values $T_{s} \sim$ 10 to 30 K, the masses fall in the range 8 to 30 $\smass$. Therefore it is of particular interest to consider collisions between clumps each a little more massive than this, since collisions involving much smaller clumps cannot produce fragmenting layers, whilst collisions between significantly larger clumps can probably be viewed as multiple collisions between smaller sub-clumps. 

In this paper we have therefore restricted the parameter space to focus on collisions between equal mass clumps, each of mass $\sim \, 75 \smass$, and we have explored how the mechanics of fragmentation and star formation changes as a function of the impact parameter of the collision.

The importance of magnetic fields in star formation is unclear. Given the low overall efficiency of star formation, it is possible that clump/clump collisions only trigger efficient star formation when the colliding clumps are threaded on the same flux tube. Under this circumstance, the critical surface density for gravitational condensation is not amplified by the collision, and once the density is high enough the gas effectively decouples from the field, so the magnetic field has little influence on the events we have simulated. Alternatively, the magnetic field may be destroyed by magnetic reconnection in certain clumps, in the manner described by \cite{lub}. 

\S\ref{sec:method} describes the simulation method and initial conditions, \S\ref{sec:bscan} discusses the results of collisions with varying impact parameter, \S\ref{sec:orient} investigates the effect of numerical noise (which is introduced by the discretisation of the gas into particles), \S\ref{sec:hires} investigates the effect of increasing the resolution of the simulation, and the overall conclusions are given in \S\ref{sec:conclude}.

\section{Method.}
\label{sec:method}

\subsection{Numerical Details.}

Hydrodynamic forces were calculated using smoothed particle hydrodynamics (`SPH', see Lucy \shortcite{luc}, Gingold \& Monaghan \shortcite{gin}, and Monaghan \shortcite{mon-sph}). SPH is a Lagrangian method which represents a continuum fluid using a set of discrete sampling points. It is widely used in astrophysical problems because of its ability to handle large density contrasts and complicated non-symmetric geometries.

The sampling points are often referred to as `particles' since they are assigned physical quantities and are moved according to the equations of motion of the fluid. A continuum quantity at arbitrary position, $A({\bf r})$, is found by performing a weighted sum using the value of $A$ at each particle, with the weights depending on the distances from ${\bf r}$. In fact, the weighting or `kernel' function has finite extent (the smoothing length $2h$) so only neighbouring particles contribute. We allow the smoothing lengths to vary freely, both in time and from particle to particle, and choose to set them such that each particle sees, typically, 45-55 neighbouring particles. We use the $\mbox{M}_{4}$ weighting kernel \cite{mon-ker}, and the standard SPH artificial viscosity \cite{mon-vis} with parameters $\alpha=1$ and $\beta=1$ (this viscosity is necessary to prevent unphysical particle penetration in colliding flows).

Gravitational forces were calculated using a Barnes-Hut tree \cite{bar}, one of the tree-code methods which exploit the fact that at sufficiently large distances (determined by some tolerance parameter) a multipole expansion can provide a good approximation to the gravitational field of a group of particles. The force calculation is reduced from an ${\cal O}(N^{2})$ process to one of ${\cal O}(N\log N)$. We choose to expand only up to the monopole term (effectively evaluating the centre of gravity of groups of particles), but use a tolerance parameter of $\theta$=0.57 \cite{sal}.

To avoid spurious two-body events, gravitational interactions between particles at small separations must be softened. The kernel-softening method was employed \cite{tur}. With this method the density profile of a particle is described by the kernel function (here, the $\mbox{M}_{4}$ kernel), and interactions are unaltered outside a given softening length $2\epsilon$. We chose to use a constant $\epsilon$, so that gravity was strictly conservative.

A leapfrog integration scheme was chosen \cite{hoc}, with the step-size allowed to vary from particle to particle. However, these step-sizes were discretised by choosing a maximum size and then successively halving; this allows the system of particles to be synchronised. A similar approach was used by Navarro \& White \shortcite{nav} (though they employed a Runge-Kutta integration scheme). Since it is necessary to know the positions and velocities of all particles when performing an acceleration calculation, these quantities were linearly interpolated as required; in this manner the positions and velocities of all particles are known at multiples of the smallest step size in use.

To ensure the code performed correctly, the following test simulations were successfully run : the free-fall collapse of a sphere of gravitating matter (a dynamic test of gravity), a 3D shocktube (a dynamic test of hydrodynamics) and the evolution of equilibrium isothermal spheres (an equilibrium test of both gravity and hydrodynamics). These tests are similar to those described in Turner et al. \shortcite{tur} (it should be noted that the code used here was written in C and was developed independently from the Fortran code described in Turner et al). In addition, the self-similar evolution of an adiabatic cloud of gas, as described by Evrard \shortcite{evr}, was also successfully modelled. This provides a dynamic test of both gravity and hydrodynamics and involves a variety of effects including rarefaction waves, freefall velocity fields and stagnation surfaces. By using appropriate boundary conditions, the similarity solution was reproduced more accurately and for much longer than either Evrard \shortcite{evr} or Theuns \shortcite{the} achieved.

\subsection{Equation of State.}

We use a barotropic equation of state, i.e. the pressure is a function of density only. If we define the isothermal sound speed $a\iso$ by
\begin{equation}
  p = a\iso^{2} \rho = \frac{\rho kT}{\bar{m}}
  \label{eqn:isoeos}
\end{equation}
(where for molecular gas $\bar{m} \sim 4\times10^{-24}$g) then $a\iso^{2}$ is given by,
\begin{equation}
  \begin{array}{ll}
    a_{0}^{2} \, , & \rho \leq \rho_{0} \, ; \\
    & \\
    \left\{ \left[ (a_{0}^{2} - a_{1}^{2}) \left( \frac{\rho}{\rho_{0}} \right) ^{-\frac{2}{3}} + a_{1}^{2} \right] ^{2} + \right. & \\
    \;\;\, \left. \left[ \left( (a_{0}^{2} - a_{1}^{2}) \left( \frac{\rho_{1}}{\rho_{0}} \right) ^{-\frac{2}{3}} + a_{1}^{2} \right) \left(\frac{\rho}{\rho_{1}} \right) ^{\gamma-1} \right] ^{2} \right\} ^{\frac{1}{2}} \, , & \rho > \rho_{0} \, ;
  \end{array}
\end{equation}
where $a_{0}\sim0.6\mbox{km s}^{-1}$, $a_{1}\sim0.2\mbox{km s}^{-1}$, $\rho_{0}= 1.2\times10^{-20}\mbox{g cm}^{-3}$, $\rho_{1} = 1\times10^{-14}\mbox{g cm}^{-3}$, and $\gamma = 7/5$. This equation of state may be described in terms of three regimes. Firstly, at low densities the gas is isothermal at $\sim$100K. Secondly, at intermediate densities the gas cools with increasing density ($T \propto \rho^{- \frac{2}{3}}$ approximately), until $T$ falls to $\sim$10K, and then the gas is approximately isothermal at $\sim$10K. Finally, at high densities the gas becomes adiabatic (because it becomes optically thick to its own cooling radiation) and $T \propto \rho ^{\gamma-1}$. A smooth join is obtained between the last two regimes by taking their geometric mean. We believe that this equation of state mimics the broad properties of protostellar gas which is optically thin at low densities (cooling radiatively with compression), before becoming optically thick (whereupon it heats up adiabatically). It should be emphasised that the adiabatic heating only becomes significant when the density has increased from its initial value of $\sim 10^{-21}\mbox{g cm}^{-3}$ to $> 10^{-13}\mbox{g cm}^{-3}$.

\subsection{Initial Conditions.}

\begin{figure*}
  \begin{minipage}{150mm}
  \plotxsize{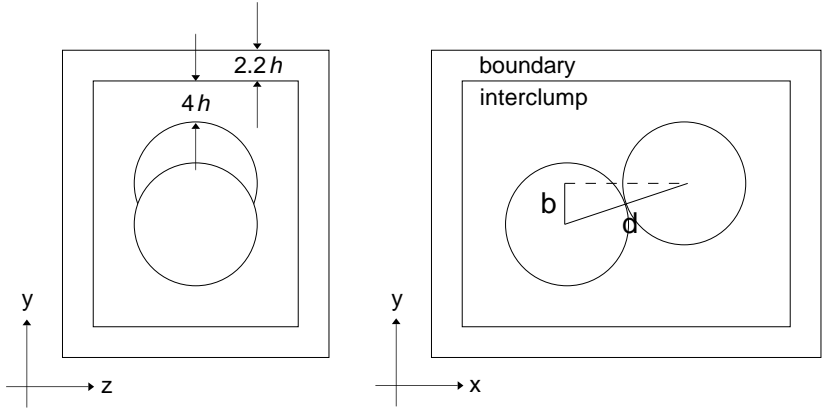}{110mm}
  \caption{Clump-clump collision initial conditions. The clumps are immersed in an interclump medium which extends $4h$ beyond the bounding box containing the clumps, and this is in turn contained within a boundary layer $2.2h$ thick. The clumps are offset by a distance `b' in the $y$ direction, and have a separation between centre-points of `d'. `d' is chosen such that the clumps are just touching. The clumps have equal, but opposite velocities in the $x$ direction.}
  \label{fig:cc-init}
  \end{minipage}
\end{figure*}

The initial conditions are intended to represent two independent clumps within a GMC, just before a collision between the clumps takes place. Details of the complicated conditions within the GMC are omitted in favour of an idealised representation, which although simplified, captures the essential aspects of the clumps and the collision process. The clumps used are individually in stable equilibrium prior to the collision (with no imposed perturbations), and so any structure which subsequently forms must result purely from the collision, the equation of state of the gas, and any low amplitude seed noise introduced by the numerical method.

The clumps are represented by spherically symmetric distributions of isothermal ideal gas in hydrostatic equilibrium. Equating the gravitational and pressure forces for a symmetric distribution of gas gives
\begin{equation}
  \frac{1}{\rho} \frac{\partial p}{\partial r} = - \frac{GM(r)}{r^{2}}.
  \label{eqn:sphericaleqm}
\end{equation}
Here $M(r) = \int 4\pi r^{2} \rho(r) dr$ (i.e. it is the mass interior to radius $r$). Using this, and writing the equation of state (\ref{eqn:isoeos}) so that $dp/dr = a\iso^{2} \rho \; d(\ln \rho)/dr$, enables (\ref{eqn:sphericaleqm}) to be re-arranged to the form
\begin{equation}
  \frac{1}{\xi ^{2}} \frac{d}{d \xi} \left( \xi^{2} \frac{d \psi}{d \xi} \right) = e^{-\psi},
\end{equation}
where the following substitutions have been made :
\begin{equation}
  \xi = \frac{(4\pi G \rho_{o})^{\frac{1}{2}} r}{a\iso}, \;\;\; \psi = \ln \left( \frac{\rho_{o}}{\rho} \right),
\end{equation}
($\rho_{o}$ is the central density of the distribution). The density and mass profiles of such spheres may now be obtained by numerical integration (see Chandresekhar \shortcite{cha} and Turner et al \shortcite{tur}). The resulting profiles are infinite in extent, so to produce a finite clump we specify the total mass of the clump and the isothermal sound speed, and we choose a value for $\xi_b$, i.e. the value of $\xi$ at the clump boundary. $\xi_b$ determines  how centrally `peaked' the density profile of the clump is.

Because the clumps are truncated isothermal spheres, an external pressure must be exerted upon them to restore equilibrium. This is done by immersing them in an initially uniform interclump medium (with density equal to that found at the edge of the clumps), and the interclump medium is in turn contained by a rigid boundary. To represent this system, three types of particles are used. `Clump' particles exert and feel forces due to both gravity and hydrodynamics, and are free to move. `Interclump' particles exert and feel hydrodynamic forces only, since they are intended to represent the rarefied regions which surround GMC clumps. These particles are also free to move. `Boundary' particles exert hydrodynamic force but remain static. All particles have equal masses. The interclump region is chosen to be sufficiently large so that the results are not corrupted by edge effects propogating inwards.

All these regions are produced from uniform, cubic particle distributions. Such a distribution is obtained by placing particles at random positions within a cubic volume, applying periodic boundary conditions, and allowing the system to evolve according to the hydrodynamic forces present. The evolution is terminated once the system has settled, i.e. the kinetic energy has decayed so that the distribution is static, and the density profile is uniform (within small scale discretisation effects). The resulting particle positions are approximately equidistant from each other, but not exactly regular (as on a grid or lattice).

The interclump and boundary regions are then produced from the uniform distribution by scaling it to obtain the required density, and joining and truncating copies of the resultant cubes to fill the required volume.

A clump is constructed by extracting a sphere from the original uniform distribution and (non-linearly) scaling it to produce the required equilibrium density profile. It is then also allowed to settle again, to remove any small-scale unbalanced forces and motions resulting from the discrete representation of the gas. This is done by allowing it to evolve according to both hydrodynamic and gravitational forces for $\sim20 t\ff$, within a spherical jacket of interclump and boundary particles.

The final distribution is produced by removing two spherical volumes from the simulation interclump distribution, and inserting both the settled clump, and a copy. The resulting initial conditions are as `quiet' as possible, and free from grids or any other regular particle alignments, and any seed noise is of very low amplitude.

Fig. \ref{fig:cc-init} depicts the initial conditions schematically and shows the co-ordinate system used. The clumps are given velocities in the $x$ direction, are aligned in the $z$ direction, and may be offset in the $y$ direction. This offset is described by the impact parameter $b$ which is the vertical distance between the centres of the clumps (`b' in the figure) divided by the clump diameter. The right hand clump (at large $x$) is always higher (at larger $y$) than the left hand clump if $b$ is non-zero. Hence $b=0.5$ requires the lower edge of the right hand clump to be at the same height as the centre of the left hand clump. The distance between the centres of the clumps (`d' in the figure) is always set to be such that the two clumps are initially just touching.

\subsection{Simulations conducted.}

All collisions were between identical clumps with temperature $T=100K$ (equating to a sound speed of $a\sim 0.6$\kms), mass $M=75\smass$, and with $\xi_b=3$. This results in a clump radius of $r\sim 0.6$pc, and central, edge and mean densities $\rho_{central}=1.1\times10^{-20}$g \pcc, $\rho_{edge}=3.8\times10^{-21}$g \pcc and $\bar{\rho}=5.6\times10^{-21}$g \pcc. The freefall time of such a clump is $t\ff\sim0.9$Myr. Each collision also took place at a relative velocity of $\sim$1.8\kms or a relative Mach number of ${\cal M}=3$ (i.e. each clump was given a velocity of $\sim$0.9\kms), and each collision was followed for a period of 1Myr.

The effect of varying the impact parameter $b$ was investigated by a series of collisions spanning $b$=0.1 to 0.7, in steps of 0.1. These collisions used 2000 particles per clump, and a gravitational softening parameter $\epsilon$=0.002 pc. 

To investigate the effect of noise introduced by representing the gas by a discrete set of particles, two additional collisions were performed at $b$=0.5, but with the clumps re-oriented to change the particle positions within them. All other parameters remained as before. 

To investigate the effect of numerical resolution, a single high-resolution collision was also conducted with impact parameter $b$=0.5, but employing 8000 particles per clump, and with a gravitational softening length of $\epsilon$=0.00127 pc.

\subsection{Presentation of the results.}
\label{subsec:presentation}

The figures showing the results of the collisions are plots of column-density in which the grey-scale is logarithmic. In addition, contours which are equally separated in log-space are overlayed on these plots.

It should be noted that the contours highlight relatively low density features. The compact condensations that form have much higher densities, but further contouring would be illegible. The high density of these protostellar condensations can be seen by examining the captions of the figures - these state the column-density range present.

The results of the collisions are summarised in tables which also give the formation time of the first compact condensation, and the values of various physical properties for the condensations formed. These properties are the radius (r$_{*}$), the maximum density (n$_{*}$), the mass (M$_{*}$), and the magnitude of the specific angular momentum about the centre of mass of the condensation (j$_{*}$). The (non-specific) magnitude of angular momentum of a condensation ($|{\bf J}_{*}|$), and the ratio of rotational energy to self gravitational energy ($\beta _{*}$) can also be calculated.

To determine these values, the constituent particles of the condensation must be found. This was done by discarding particles below a density threshold of $\rho =6.8\times10^{-16}$g \pcc, and then associating each particle with its densest neighbour. The peaks of condensations may then be found by searching for particles which have themselves as densest neighbours, and the set of particles belonging to the condensation are those which point towards the peak directly, or indirectly through other particles. The values of the properties listed above may then be calculated easily.

The centres of mass may also be calculated, and performing this process over a time sequence of particle data produces the paths taken by the condensations. Figures of this type are used to illustrate the paths of objects produced after fragmentation.

Since the collisions take place with clumps offset in the $y$-direction, and with velocities in the $x$-direction, the only non-zero component of angular momentum is the $z$-component, $L_{z}$. Since $L_{z}$ should be conserved, it can be used as an indicator of the accuracy of the code - the change in $L_{z}$ for the entire system (i.e. summed across all particles within both clumps) should be small. The maximum percentage change for each simulation is given in the summary tables.

\section{Varying the impact parameter.}
\label{sec:bscan}

As the impact parameter $b$ is altered, the mechanism by which protostellar discs are produced is changed. We concentrate on examining results from the $b$=0.3, 0.5, and 0.7 collisions. Results from all the collisions are summarised in Table \ref{table:bscan}, and discussed in \S\ref{subsec:dependence}.

The clumps were each represented by isothermal spheres of 2000 particles. The total number of particles used varied from 33755 for the $b$=0.1 collision (2000 per clump, 16519 representing the interclump, and 13236 forming the boundary) to 38896 for the $b$=0.7 collision (2000 per clump, 20284 representing the interclump, and 14612 forming the boundary).

\begin{table*}
\begin{minipage}{150mm}
\caption{Summary of the standard collisions (see \S\ref{sec:bscan}).}
\label{table:bscan}
\begin{tabular}{c|ccccccccc}
$b$ & $t_{*}$ & $r_{*}$ & $n_{*}$ & $M_{*}$ & $j_{*}$ & $\Delta L_{z}$\\
\footnotesize (/dia) & \footnotesize (Myr) & \footnotesize (AU) & \footnotesize (10$^{11}$cm$^{-3}$) & \footnotesize ($\smass$) & \footnotesize (10$^{24}$cm$^{2}$s$^{-1}$) & \footnotesize (\%) \\ \hline
& & & & & & & & \\
0.1 & 0.63 & 1330 & 16 & 60 & 2.1 & -2 \\
& \multicolumn{7}{l}{\footnotesize One central object. \normalsize} \\
& & & & & & & & \\
0.2 & 0.62 & 1030 & 12 & 60 & 2.2 & -2 \\
& \multicolumn{7}{l}{\footnotesize Six objects (2-60$\smass$), one dominant plus two smaller objects formed by rotational instability. \normalsize} \\
& & & & & & & & \\
0.3 & 0.64 & 880,1130 & 3.0,5.3 & 21,40 & 1.3,2.0 & -2 \\
& \multicolumn{7}{l}{\footnotesize Three objects (3-40$\smass$), two dominant. Secondary formed by rotational instability from primary. \normalsize}\\
& & & & & & & & \\
0.4 & 0.68 & 580,860 & 4.0,2.7 & 13,27 & 0.5,1.5 & -1 \\
& \multicolumn{7}{l}{\footnotesize Six objects (2-27$\smass$), two dominant. Secondary formed by lumpy accretion flow onto primary. \normalsize} \\
& & & & & & & & \\
0.5 & 0.78 & 890,920 & 1.2,1.5 & 17,17 & 1.3,1.3 & -1 \\
& \multicolumn{7}{l}{\footnotesize Two objects, each formed by multiple fragment mergers, falling towards each other. \normalsize} \\
& & & & & & & & \\
0.6 & 0.89 & 840,830 & 0.5,0.8 & 11,11 & 1.2,1.1 & -1 \\
& \multicolumn{7}{l}{\footnotesize Two objects, well separated. \normalsize} \\
& & & & & & & & \\
0.7 & 0.94 & 680,650 & 0.4,0.4 & 7,7 & 0.8,0.8 & -1 \\
& \multicolumn{7}{l}{\footnotesize Two objects, well separated. \normalsize} \\
& & & & & & & & \\
\end{tabular}
$r_{*}$, $M_{*}$ and $j_{*}$ are determined using particles above a density threshold of $n=1.7\times10^{8}$ cm$^{-3}$. $t_{*}$ is the earliest time 2$\smass$ of material forms a compact condensation. $n_{*}$ is the maximum density of a protostar at the end of the simulation.
\end{minipage}
\end{table*}

\subsection{$b$=0.3}

\begin{figure*}
  \begin{minipage}{150mm}
    \plottwo{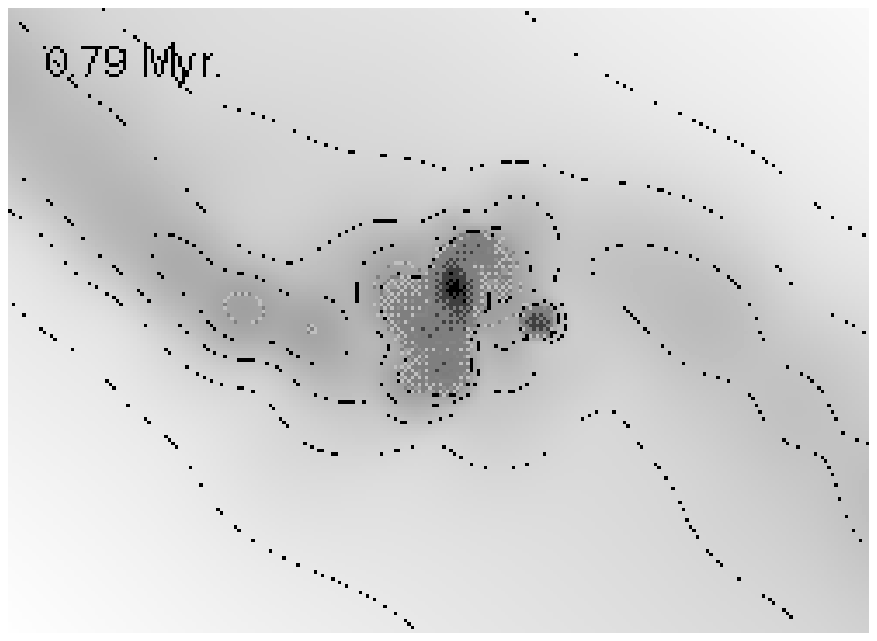}{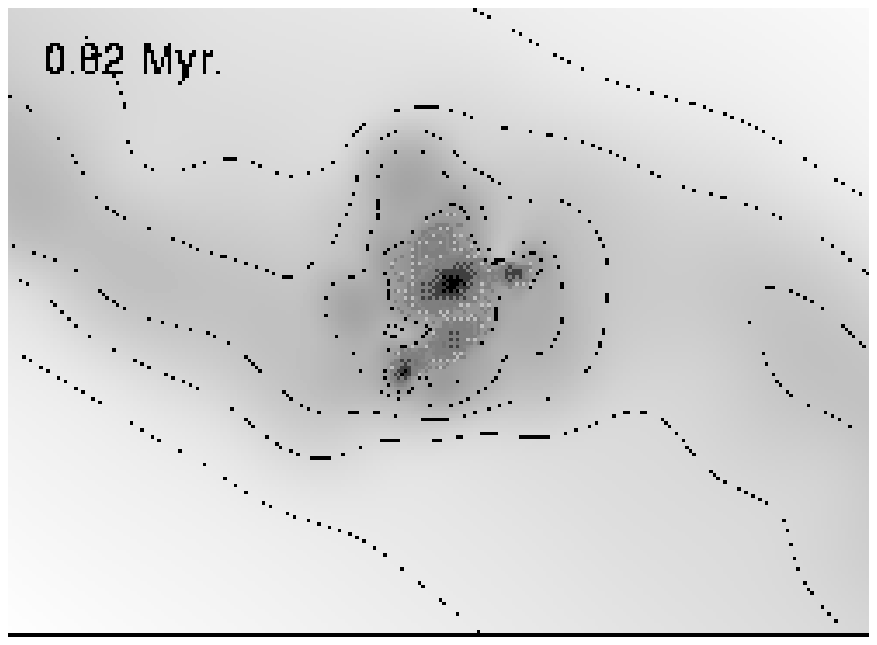}\\
    \vspace{6pt}
    \plottwo{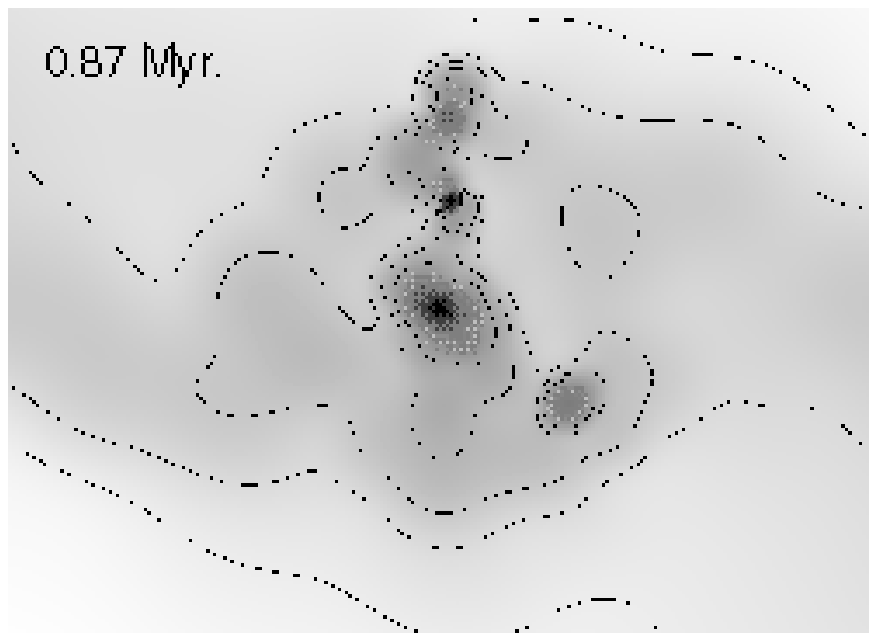}{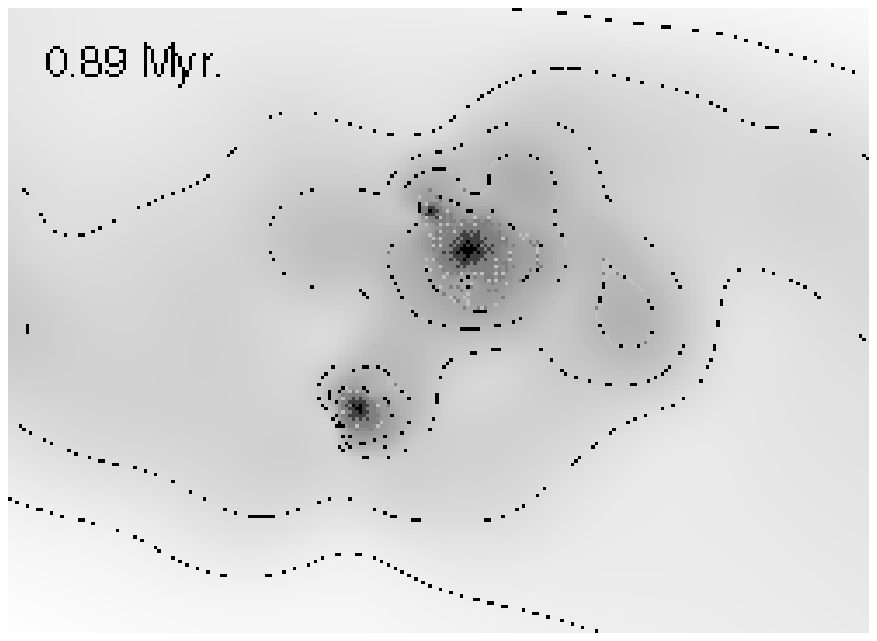}\\
    \vspace{6pt}
    \plottwo{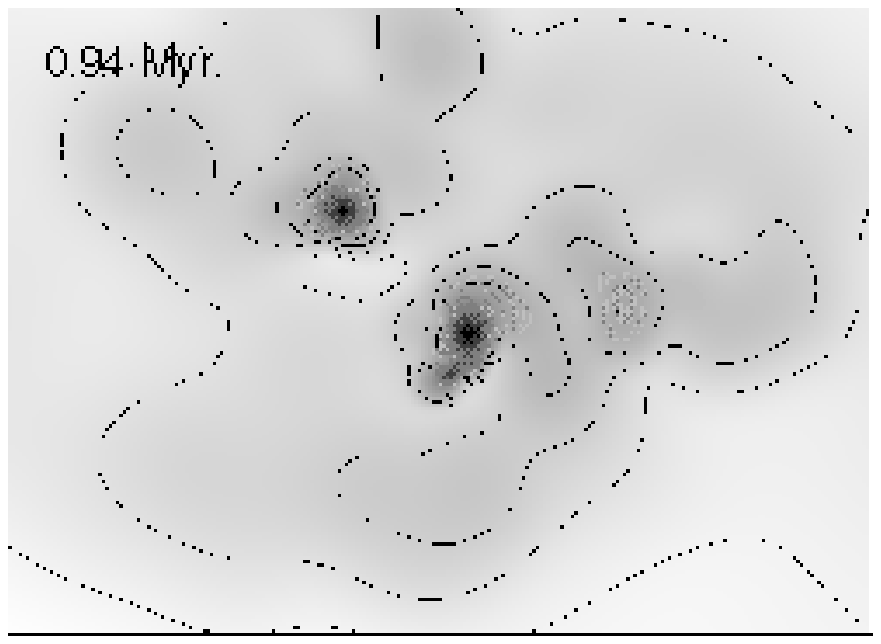}{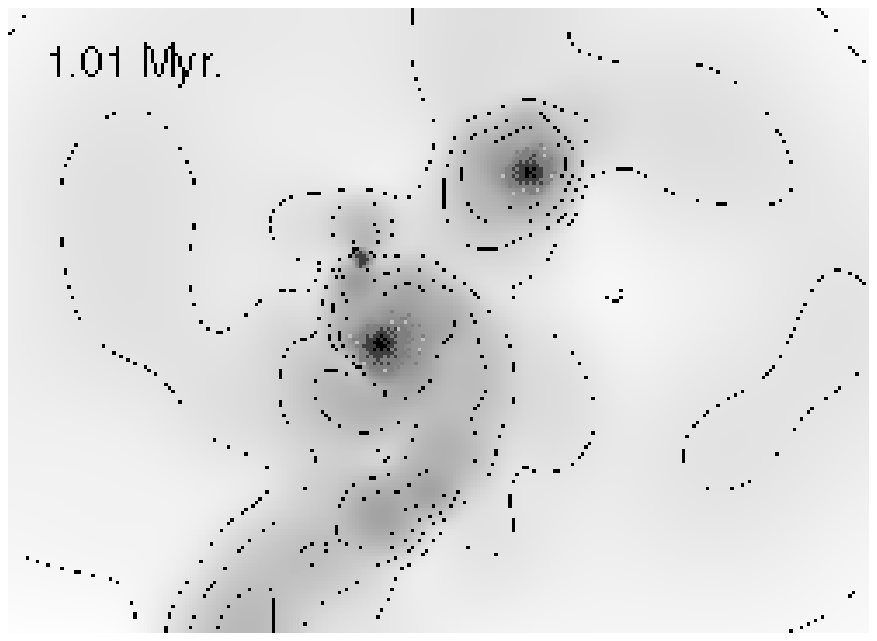}\\
  \caption{$b$=0.3 collision evolution. Accretion onto a central condensation is followed by rotational instabilities. [$x$=(-0.08,0.08)pc, $y$=(-0.06,0.06)pc, contours at $\logn$ = (22.4, 22.7, 23.1, 23.4, 23.7), min=22.4, max=27.2.]}
  \label{fig:cc-b0.3evolve}
  \end{minipage}
\end{figure*}

The evolution of the system was dominated by accretion onto a central condensation and its subsequent rotational instability (Fig. \ref{fig:cc-b0.3evolve}). 

A single rotating spherical condensation rapidly formed at the collision centre, within a shock which was thin in the $x$-$y$ plane, and somewhat extended in the $z$ direction. The shock subsequently became more spindle-shaped (less extended in the $z$ direction), and the condensation more disc-like. The smooth accretion flow along the shock caused the condensation to grow in size and mass, and increased its specific angular momentum causing it to became increasingly rotationally unstable. Material was then thrown off by the disc via weak spiral arms, and a surrounding diffuse envelope was produced out of which a distinct secondary condensed. This subsequently orbited the primary and grew in mass by `mopping-up' the surrounding diffuse material, and by periodically intercepting the accretion flow. The final state consisted of a binary system of two protostellar discs, with a primary of 40$\smass$ and 1100 AU radius, and a secondary of 21$\smass$ and 880 AU radius. The final separation was 8820 AU, and the secondary completed two orbits of the primary.

\begin{figure}
  \plotxsize{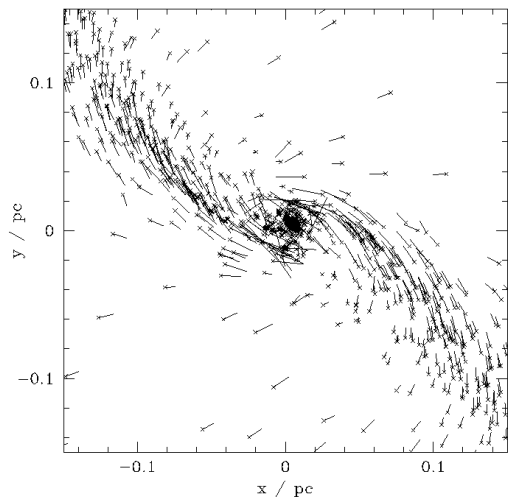}{70mm}
  \caption{$b$=0.3 collision. Offset accretion flow onto the central condensation at $t$=0.79 Myr. Crosses mark particle positions and lines are velocity vectors. Vectors are only plotted for particles in regions below a threshold density ($\sim 1\times10^{-17}$g \pcc), leaving the central condensation particle positions visible.}
  \label{fig:cc11-acc}
\end{figure}

\begin{figure*}
  \begin{minipage}{150mm}
  \plottwo{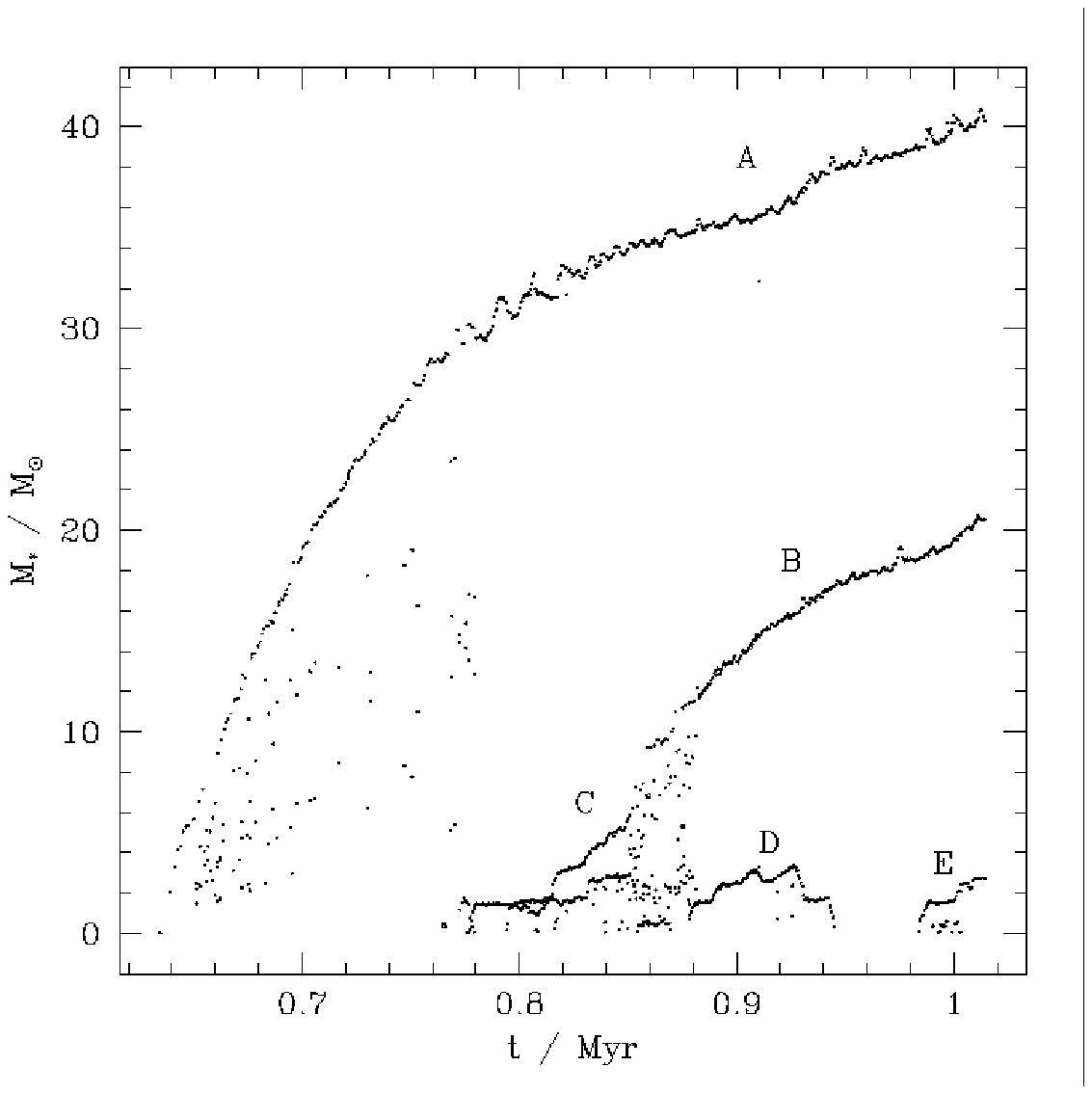}{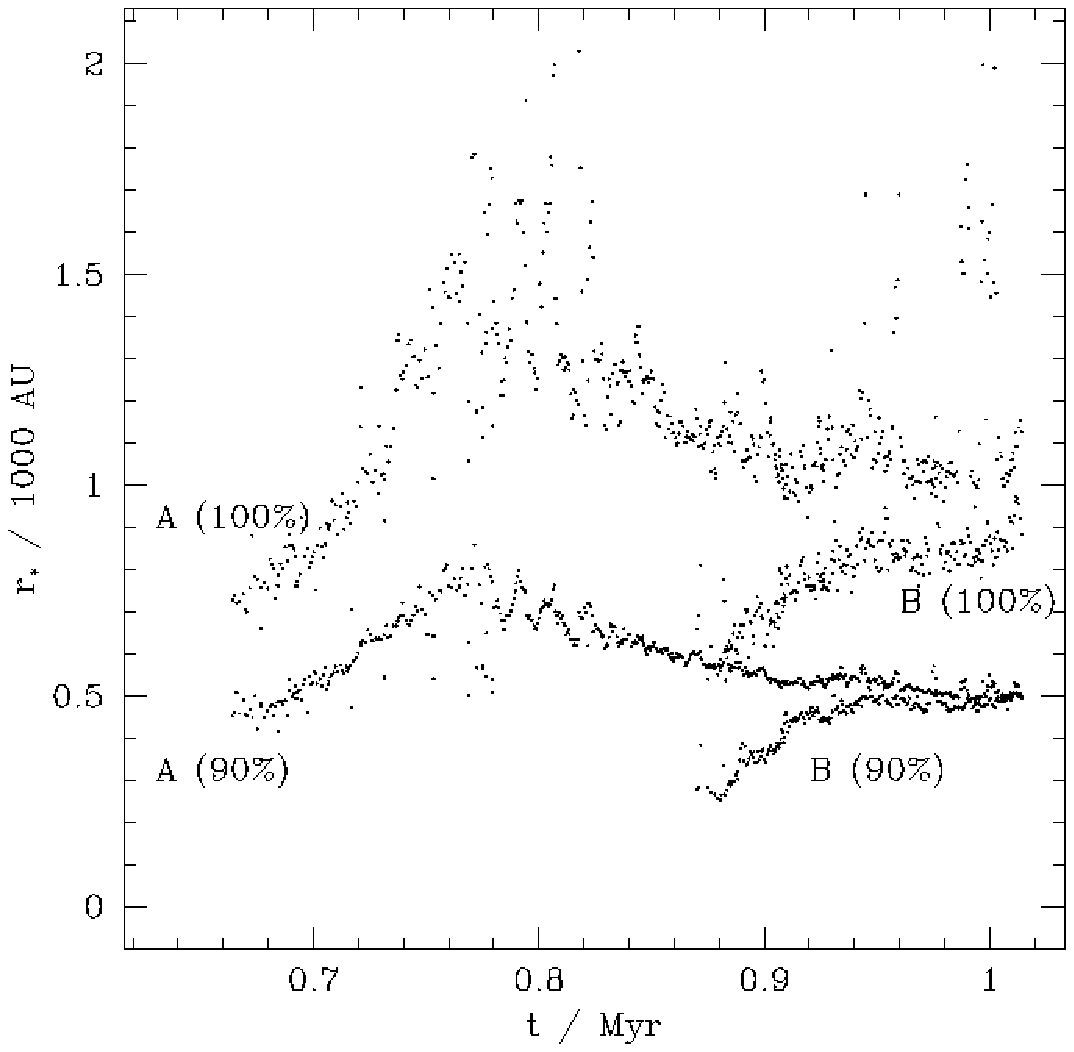}\\
  \plottwo{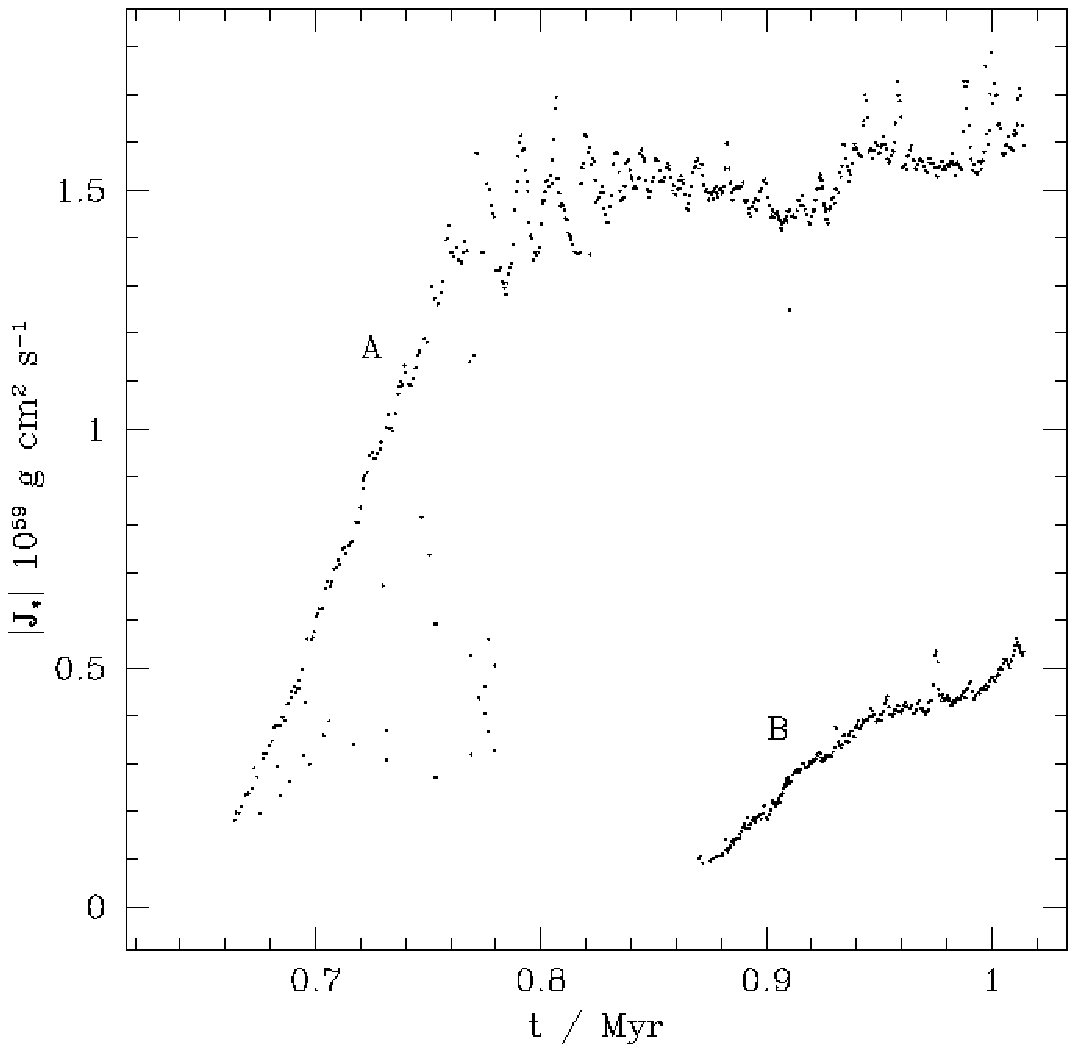}{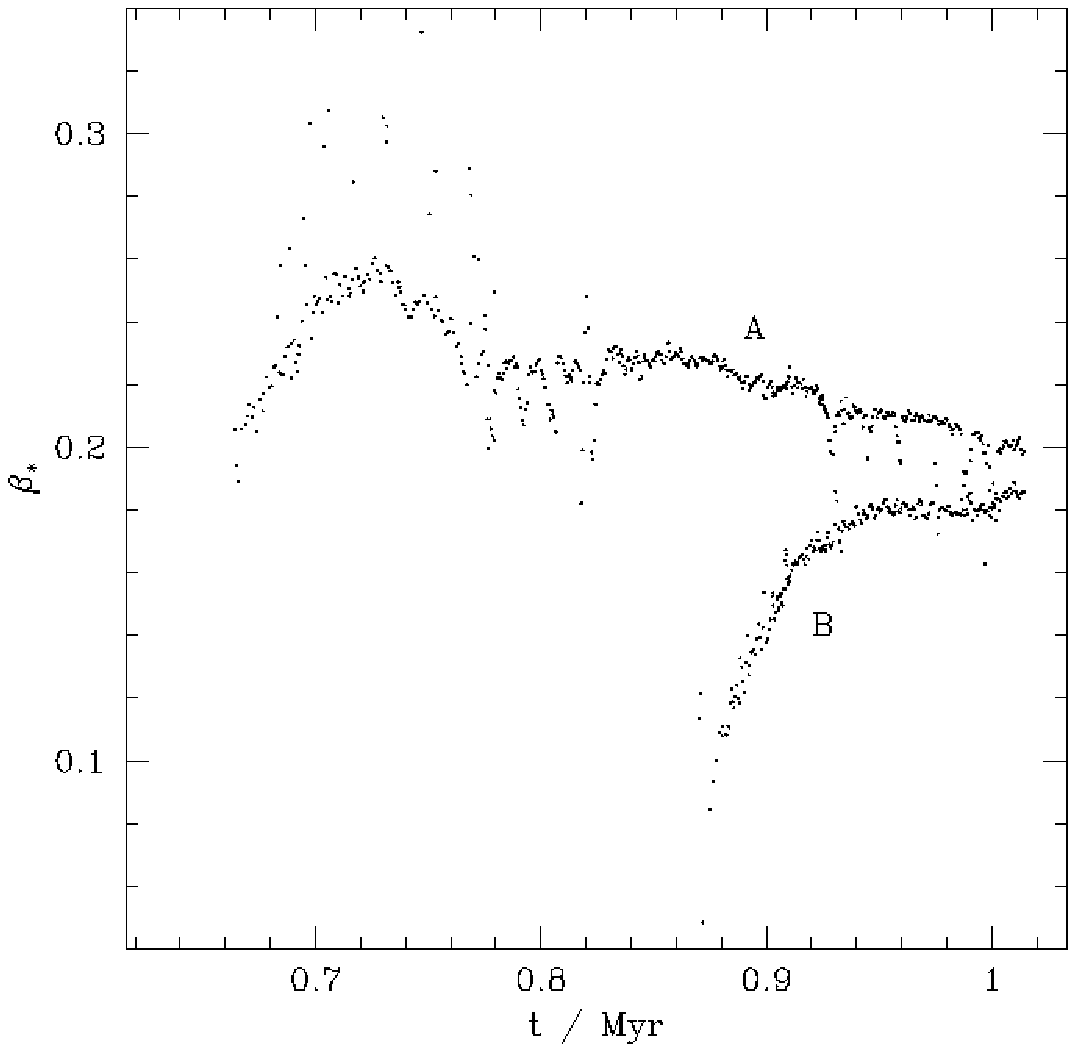}\\
  \caption{$b$=0.3 collision. Evolution of condensation physical properties : mass M$_{*}$, 100\% and 90\% radii r$_{*}$, angular momentum about the condensation centre of mass $|{\bf J}_{*}|$, and the ratio of rotational to self gravitational energy $\beta _{*}$. A minimum mass threshold of 10$\smass$ was applied to all except the mass plot. All graphs show the evolution of the primary (A), and the secondary (B). The mass evolution graph also shows the two fragments which merge to produce the secondary (C), a transient fragment thrown off and subsequently disrupted by the primary (D) and another fragment produced by the primary (E).}
  \label{fig:cc11-phys}
  \end{minipage}
\end{figure*}

The evolution of various physical properties of the condensations is shown in Fig. \ref{fig:cc11-phys}, and is discussed below. The subscripts `1' and `2' will be used here to denote properties for the primary and secondary condensations respectively. 

Initially the accretion flow within the shock-bounded spindle is focussed onto the central primary object, but a small offset rapidly develops, causing material streaming in from the upper-left to meet the lower edge of the condensation, and material from the lower-right to meet the upper edge of the condensation (Fig. \ref{fig:cc11-acc}). The finite impact parameter of the collision causes the spindle to rotate about the angular momentum axis of the collision (the $z$-axis), and the accretion flow within the spindle evolves further and further away from the initial linear configuration, becoming progressively more s-shaped and offset from the collision centre. Thus the angular momentum introduced to the central condensation is steadily increased, and the condensation is spun-up ($|{\bf J}_{1}|$ and M$_{1}$ increase rapidly), causing it to become unstable ($\beta _{1}$ increases). This can be seen in Fig. \ref{fig:cc11-phys} at $t<$0.76 Myr.

Once $\beta _{1} \sim 0.25$ and $|{\bf J}_{1}| \sim 1.4\times10^{59}$ g cm$^{2}$ s$^{-1}$, material is shed via rotational instabilities (at $t$=0.76 Myr). Shedding and re-accretion cause oscillations in M$_{1}$ and $|{\bf J}_{1}|$ until two fragments are permanently shed ($t$$\sim$0.77 Myr). It is worth noting that the instability does not just produce a simple spiral-arm structure. The process is more irregular, as weak spiral arms are repeatedly formed and merge with diffuse material ejected earlier, and with material from the accretion flow. The two fragments produced are visible as the two traces below the label `C' in the mass evolution plot of Fig. \ref{fig:cc11-phys}. These subsequently merge to produce the secondary (at $t$=0.86 Myr, the beginning of trace `B' in Fig. \ref{fig:cc11-phys}).

The increasing offset of the accretion flow causes incoming material to swirl in with a large angular momentum relative to the discs. As the secondary orbits the primary it intercepts the flow and gains mass and angular momentum from it, so that M$_{2}$ and $|{\bf J}_{2}|$ increase, in addition some of the diffuse material around the primary is also collected. Meanwhile, the primary is both accreting and occasionally shedding material from the edge of the disc (thus removing angular momentum, e.g. objects `D' and `E' in the mass evolution graph in Fig. \ref{fig:cc11-phys}). M$_{1}$ increases slowly whilst $|{\bf J}_{1}|$ remains approximately constant. Thus $\beta _{1}$ (and hence the amount of rotational support) is decreased and the disc shrinks slightly (the 90\% mass radius falls from 700 AU to 500 AU), whilst the peak density rises (continuing an almost linear trend since the formation of the primary).

\begin{figure*}
  \begin{minipage}{150mm}
  \plottwo{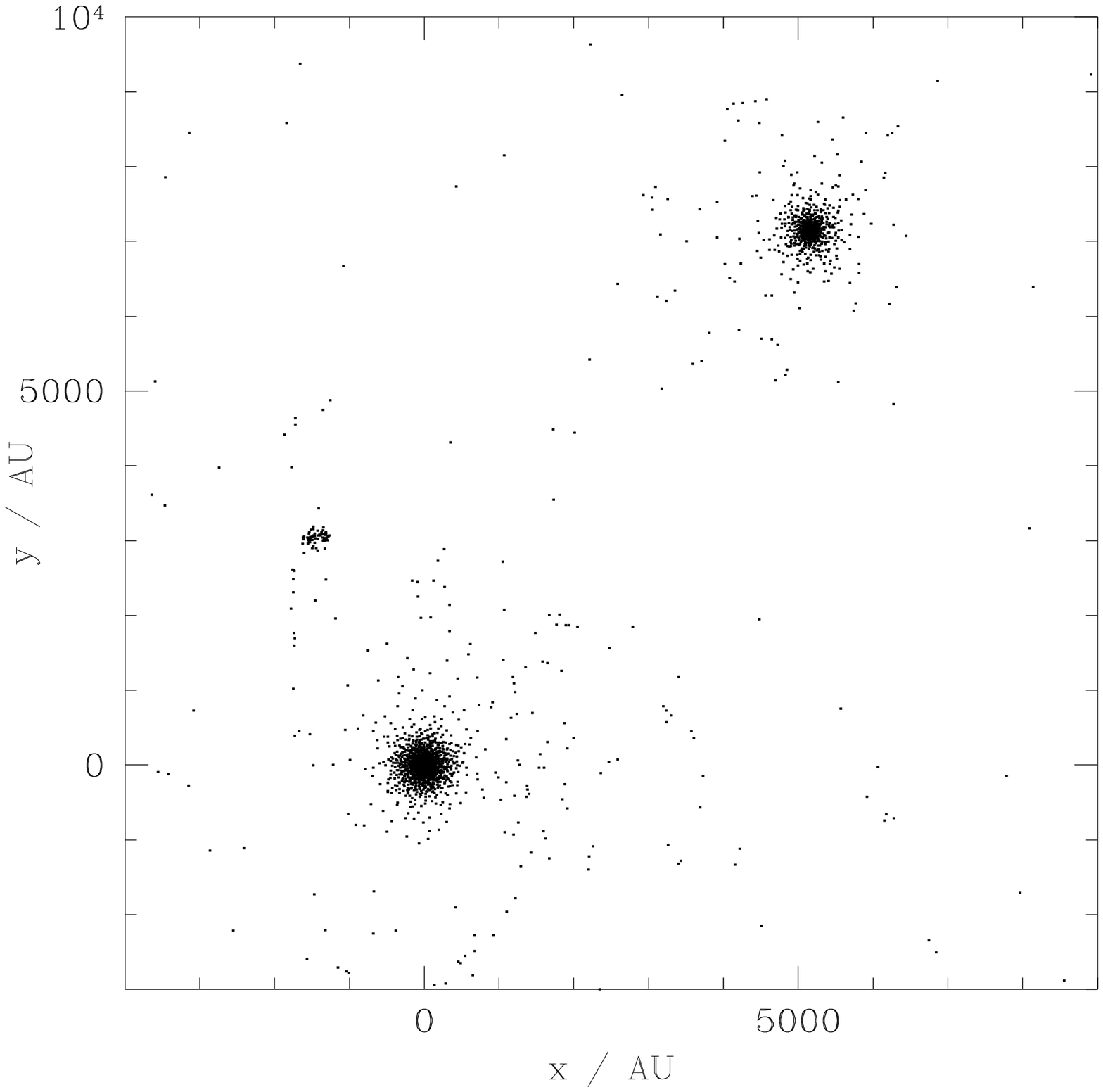}{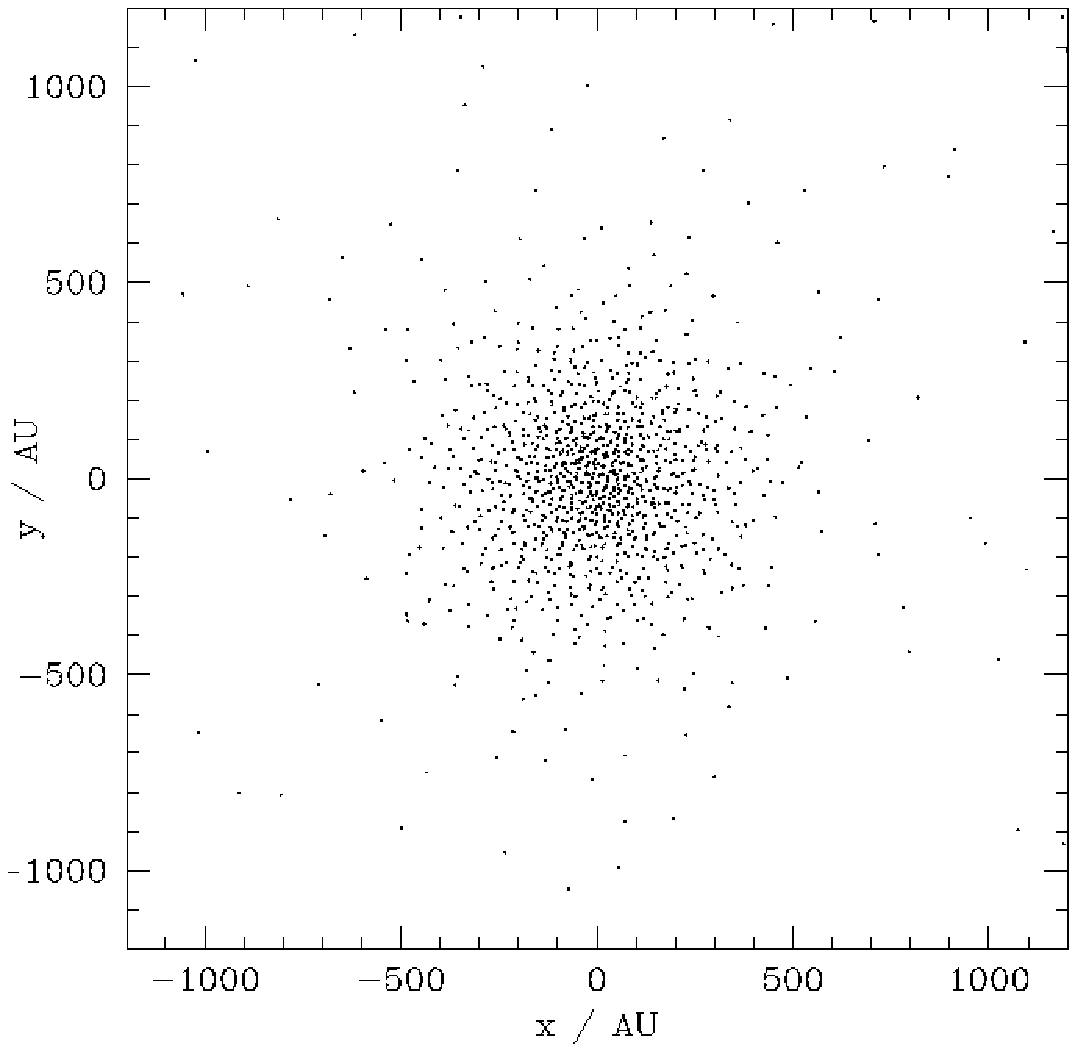}\\
  \plottwo{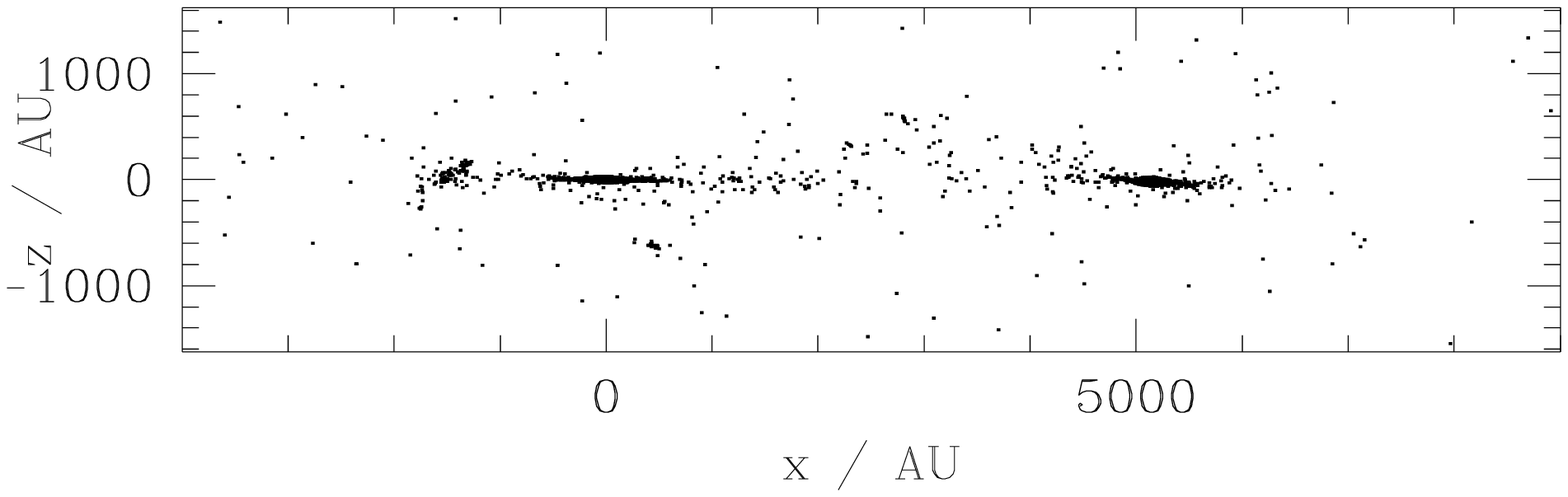}{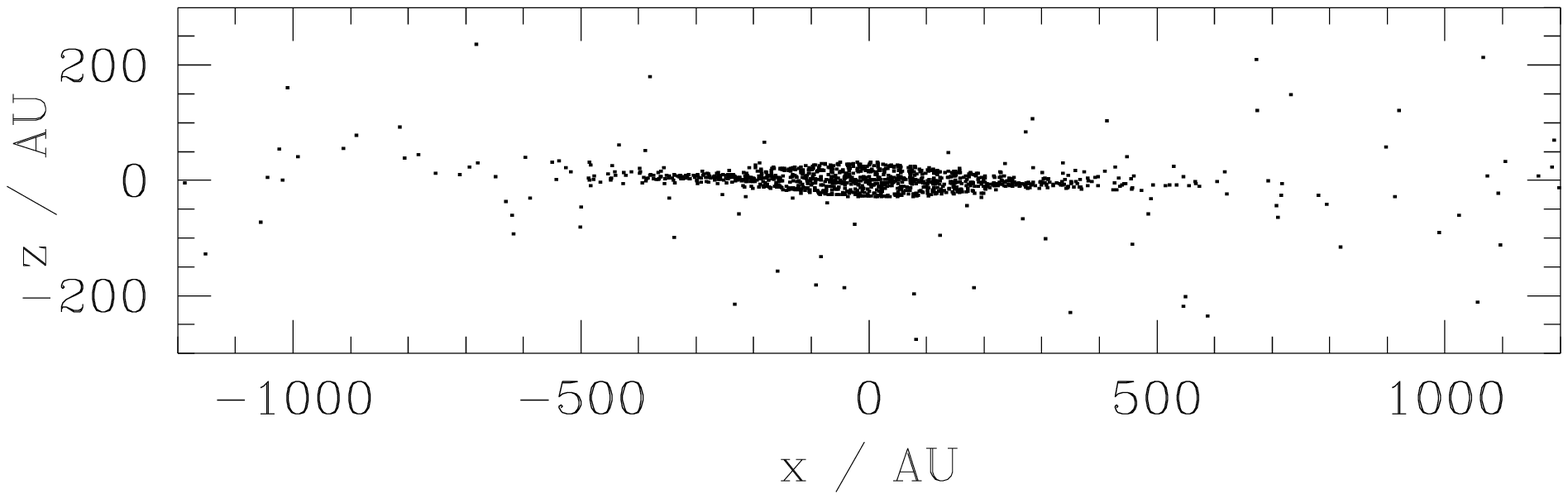}\\
  \caption{$b$=0.3 collision. Protostellar discs at the end of the simulation (particle positions). Co-ordinates are centred on the centre of mass of the primary. Primary and secondary (left), and primary only (right).}
  \label{fig:cc11-discs}
  \end{minipage}
\end{figure*}

Fig. \ref{fig:cc11-discs} demonstrates that the protostellar objects formed are discs, and that they are co-planar. The detailed internal structure of the discs is poorly modelled due to gravity softening, and excessive shear viscosity, and so is not examined (our main aim here is with the dynamics which form the discs rather than the subsequent evolution).

\subsection{$b=0.5$.}

\begin{figure*}
  \begin{minipage}{150mm}
    \plottwo{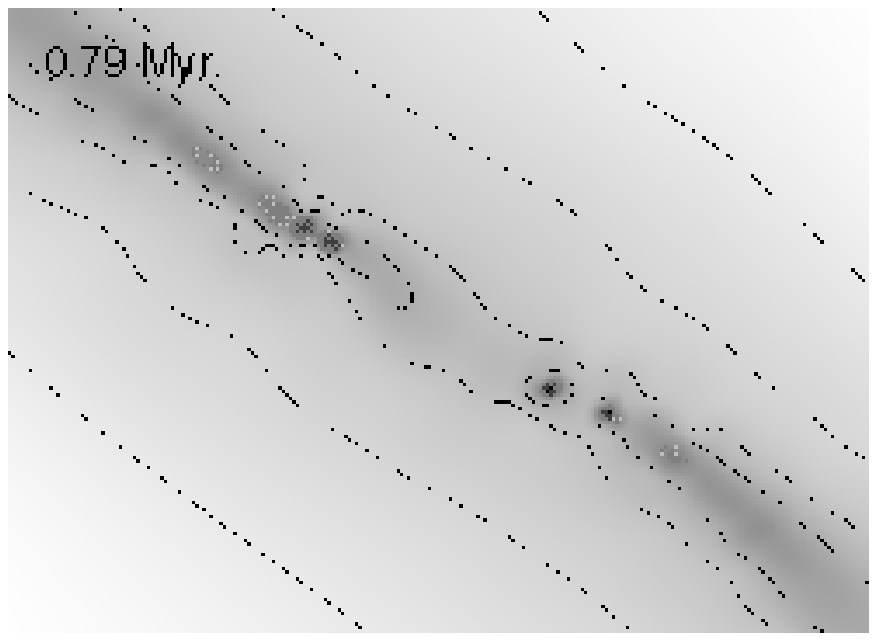}{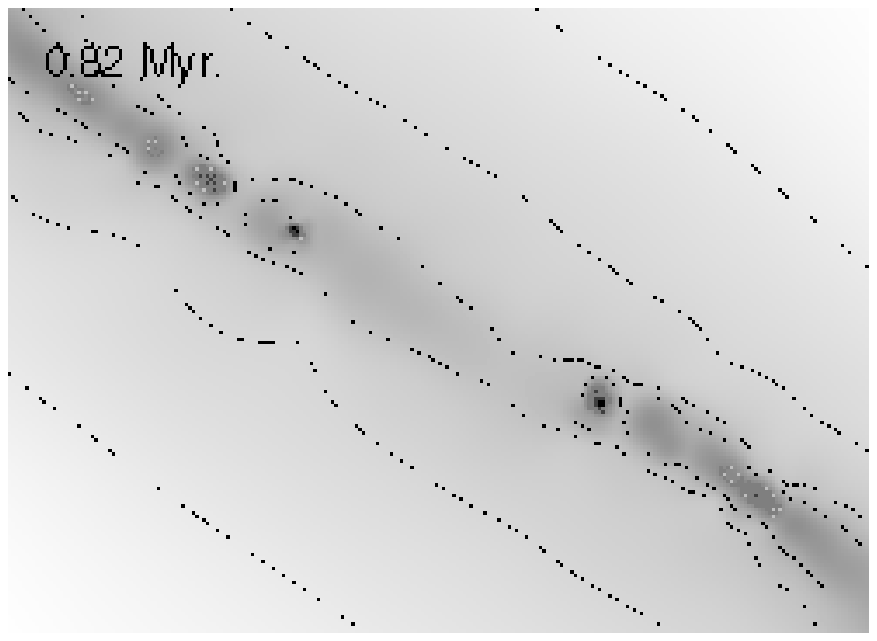}\\
    \vspace{6pt}
    \plottwo{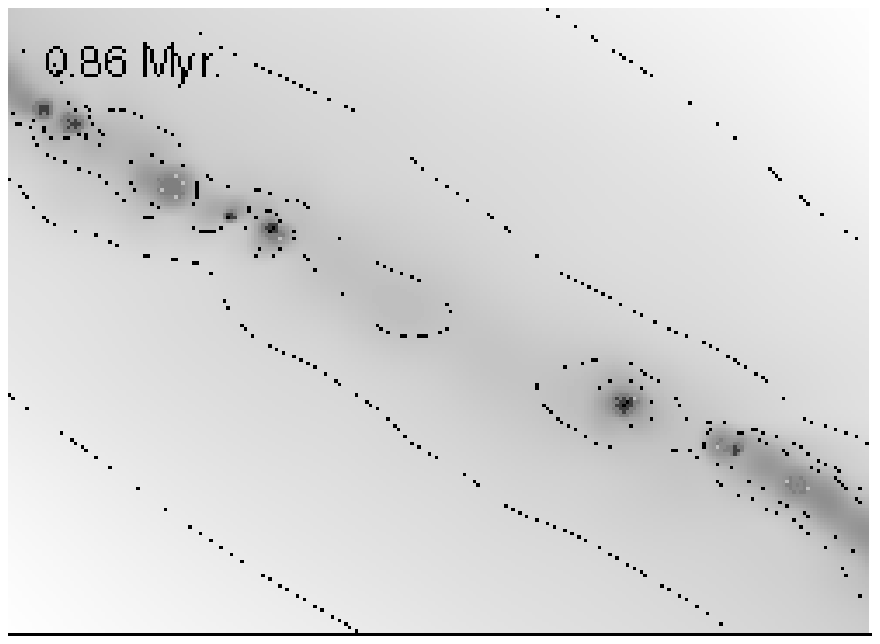}{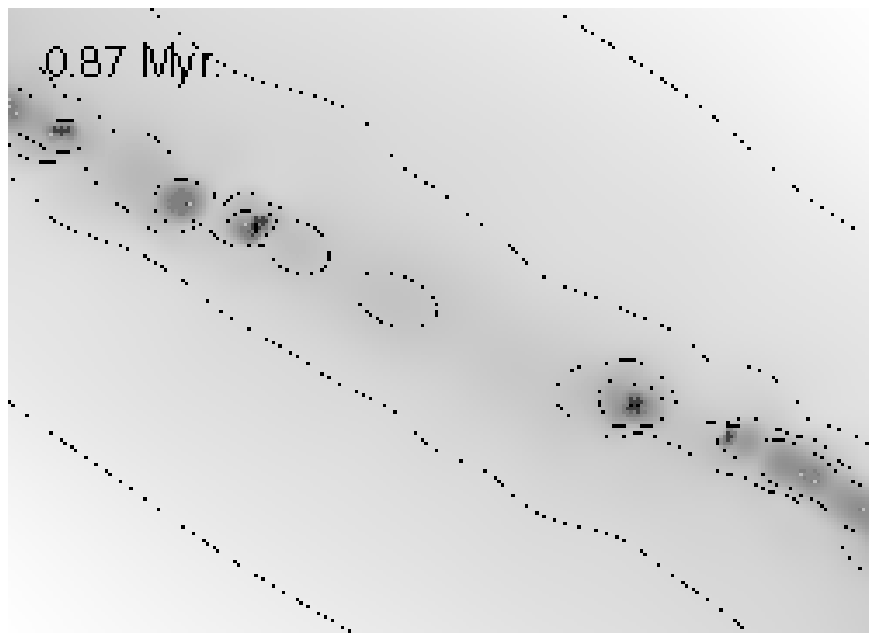}\\
    \vspace{6pt}
    \plottwo{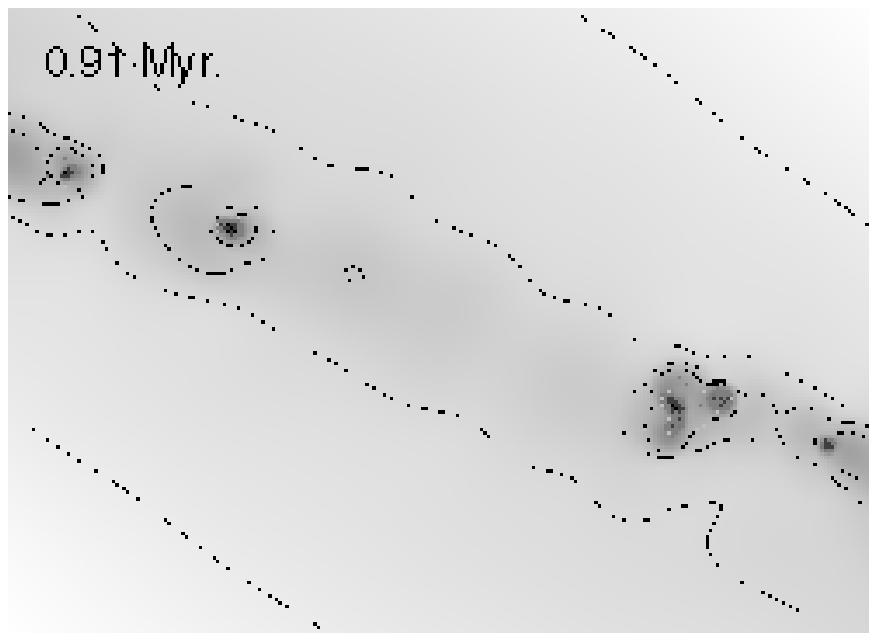}{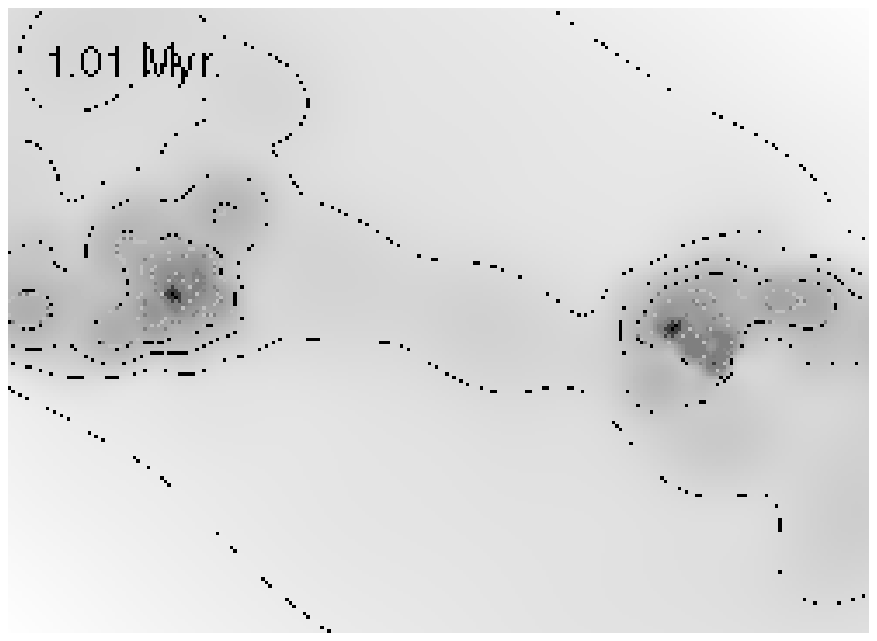}\\
  \caption{$b$=0.5 collision evolution. Fragmentation in the shock is followed by merger between fragments. [$x$=(-0.20,0.20)pc, $y$=(-0.15,0.15)pc, contours at $\logn$ = (21.9, 22.3, 22.6, 22.9), min=21.6, max=26.4.]}
  \label{fig:cc-b0.5evolve}
  \end{minipage}
\end{figure*}

The evolution of the system was dominated by fragmentation and subsequent merger within the shock (Fig. \ref{fig:cc-b0.5evolve}). Initially, a single shock filament extending diagonally in the $x$-$y$ plane was produced at the collision interface of the clumps. At about $t$=0.79 Myr, four fragments appeared in the shock in two close pairs, one above and to the left of the centre, and the other below and to the right. Each pair subsequently merged and accreted any further fragments which developed to produce a final state of two well defined protostellar condensations of similar size ($\sim$800 AU radius) and mass (17$\smass$), with a separation of 0.23 pc. These discs were falling towards each other along the shock at the end of the simulation, and would probably capture each other at periastron.

\begin{figure*}
  \begin{minipage}{150mm}
  \plottwo{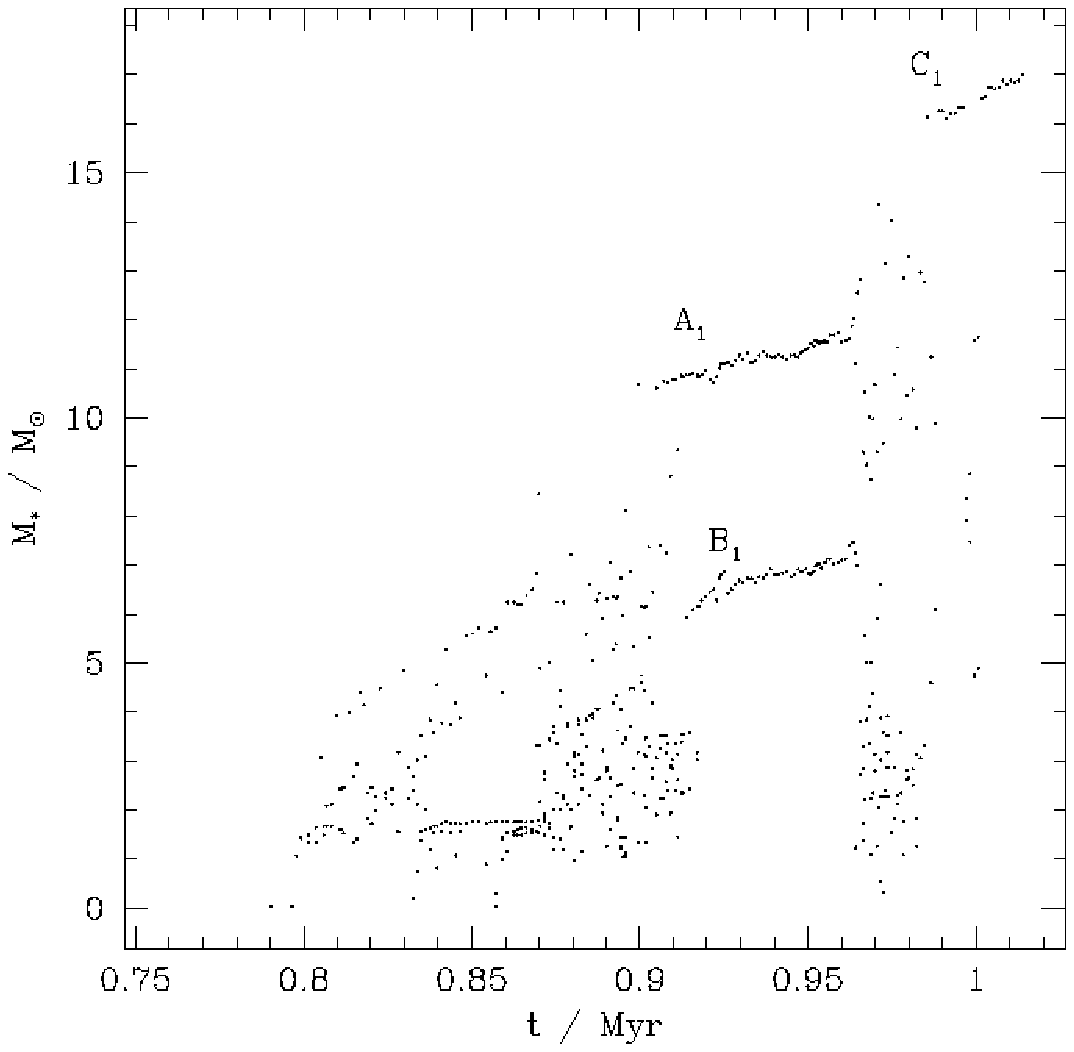} {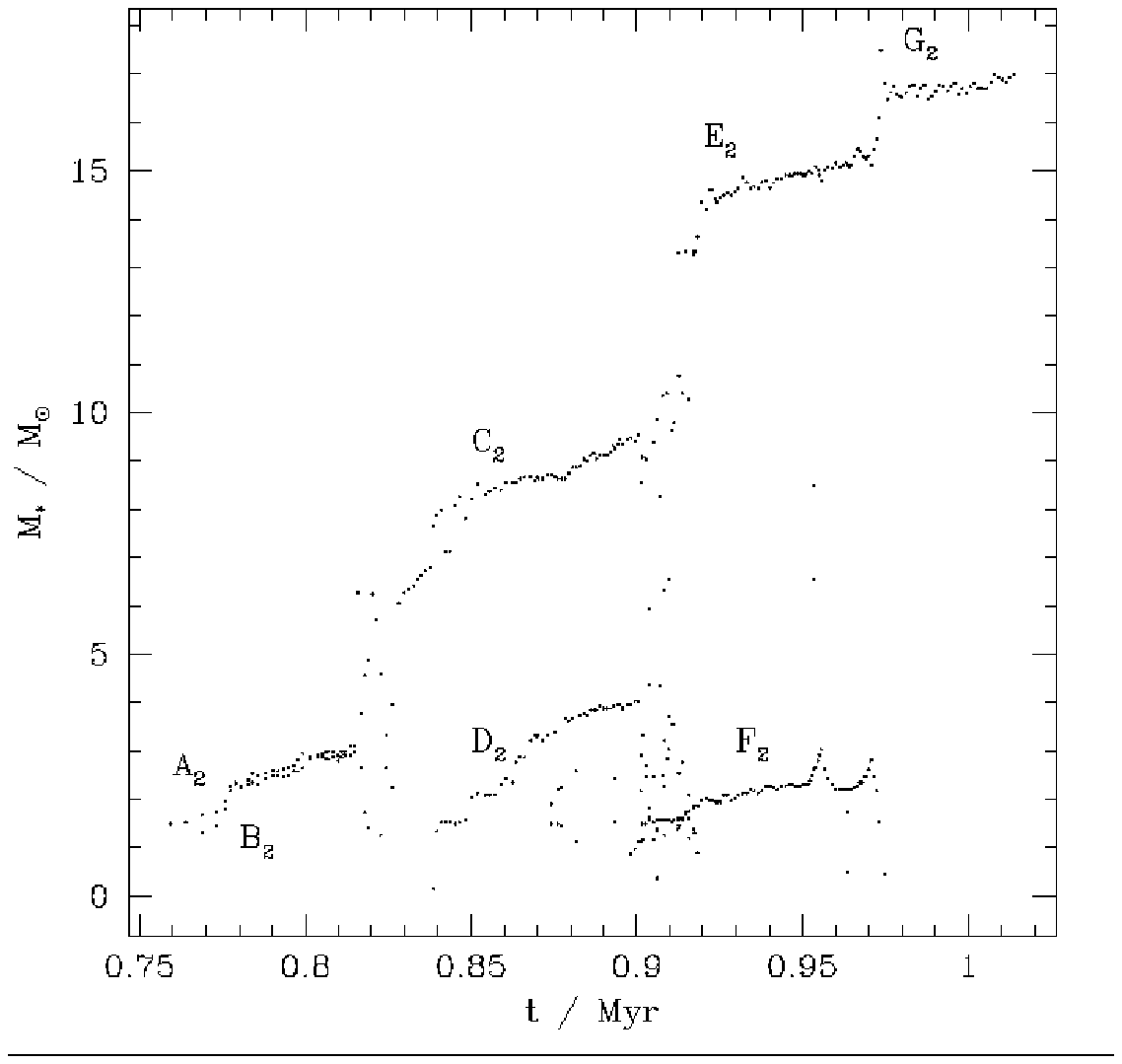} \\
  \plottwo{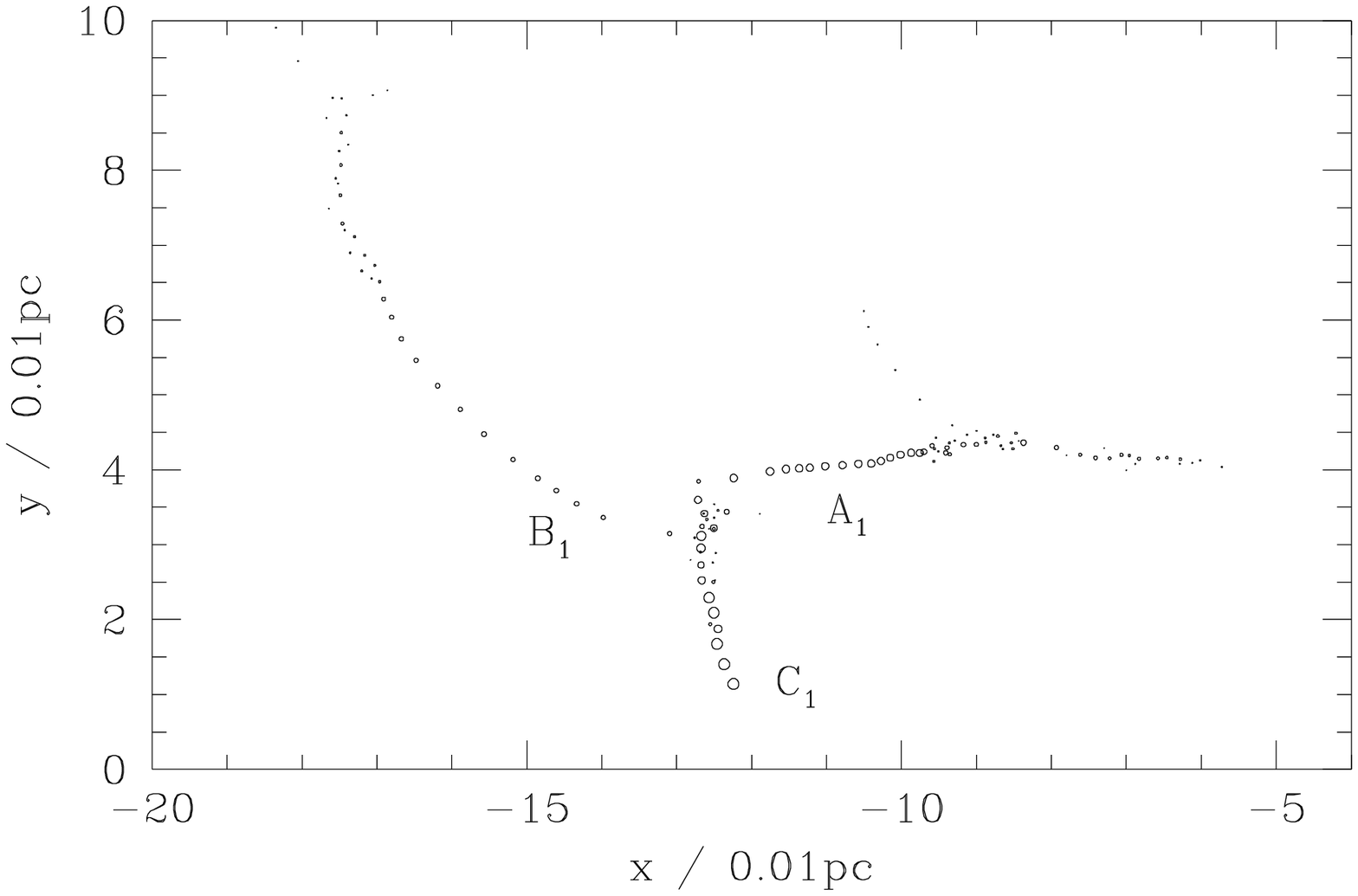}{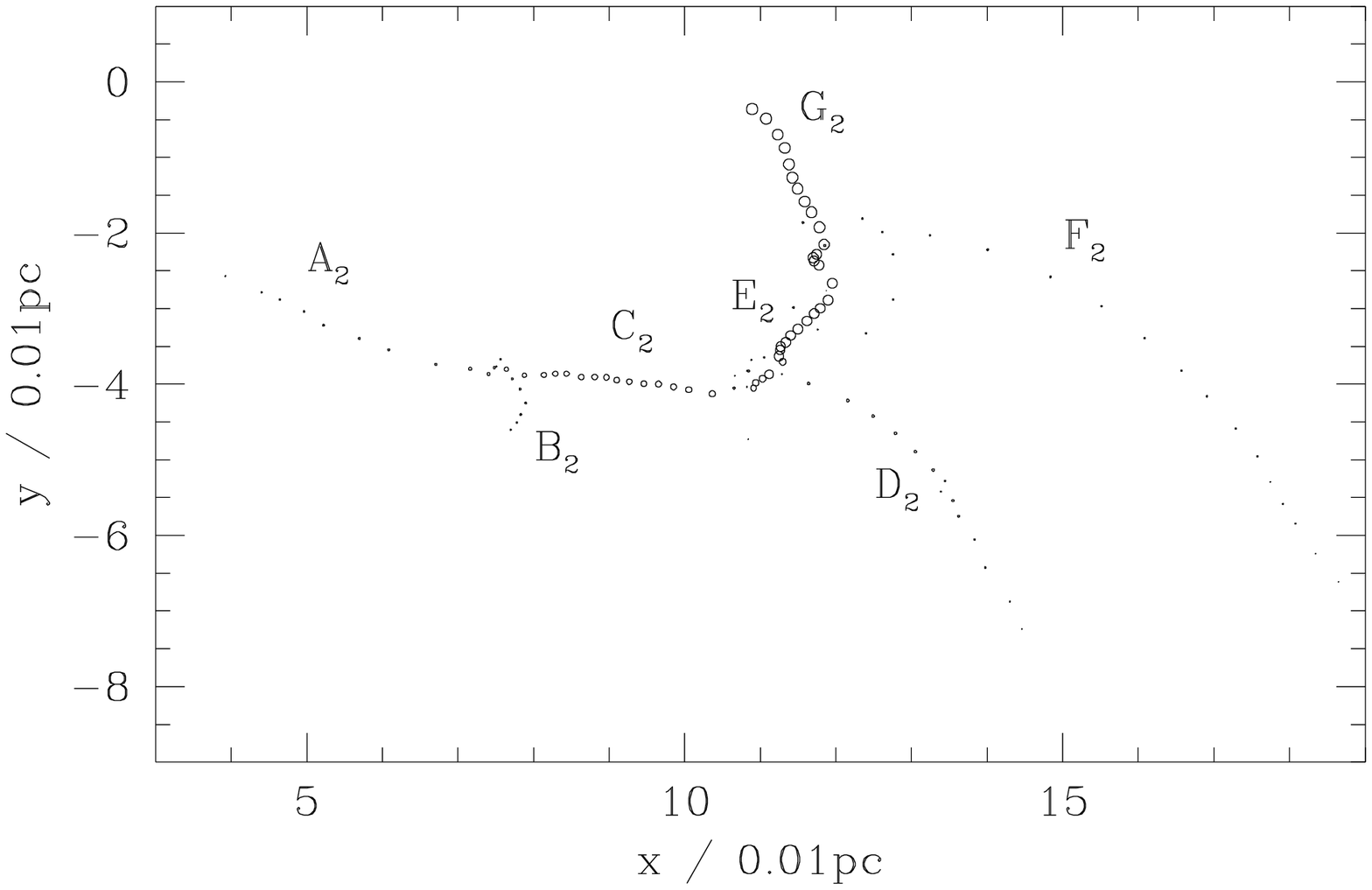} \\
  \caption{$b$=0.5 collision. Evolution of condensation masses and  positions (position markers scale linearly with mass). Condensations appearing to the left and above the centre of collision (left) are separated from those appearing to the right and below (right). Each fragment is labelled so that the corresponding masses and positions are apparent.}
  \label{fig:cc09-phys}
  \end{minipage}
\end{figure*}

Figure \ref{fig:cc09-phys} illustrates the paths and mass evolutions of the fragments throughout the simulation, with corresponding positions and masses labelled. Fragments within the upper-left region are shown in the left-hand plots, and fragments within the lower-right region are shown in the right-hand plots. The evolution of both regions is broadly similar, with fragmentation initially producing objects of 1$\smass$ and lower (though these are equivalent to about 20-30 particles, and so are poorly sampled). Merger events are clearly visible, for example in the lower-right region fragments A$_{2}$ and B$_{2}$ merge to produce C$_{2}$ at 0.82 Myr, D$_{2}$ forms at about the same time and merges with C$_{2}$ to produce E$_{2}$ at $\sim$0.91 Myr, and finally fragment F$_{2}$ appears and merges with E$_{2}$ (at 0.97 Myr) to produce the final disc G$_{2}$. These objects are also visible at the right-hand side of the column density plots of Fig. \ref{fig:cc-b0.5evolve}.

The paths taken by the fragments were affected by the shock morphology. The shock initially formed at the collision centre, and then propagated outwards as more of the clumps came into contact. Thus, the sequence of new fragments, i.e. those not produced by merger, similarly propagated outwards. At the lower-right region this sequence of fragments is A$_{2}$, B$_{2}$, D$_{2}$, and F$_{2}$.

\subsection{$b=0.7$.}

\begin{figure*}
  \begin{minipage}{150mm}
    \plottwo{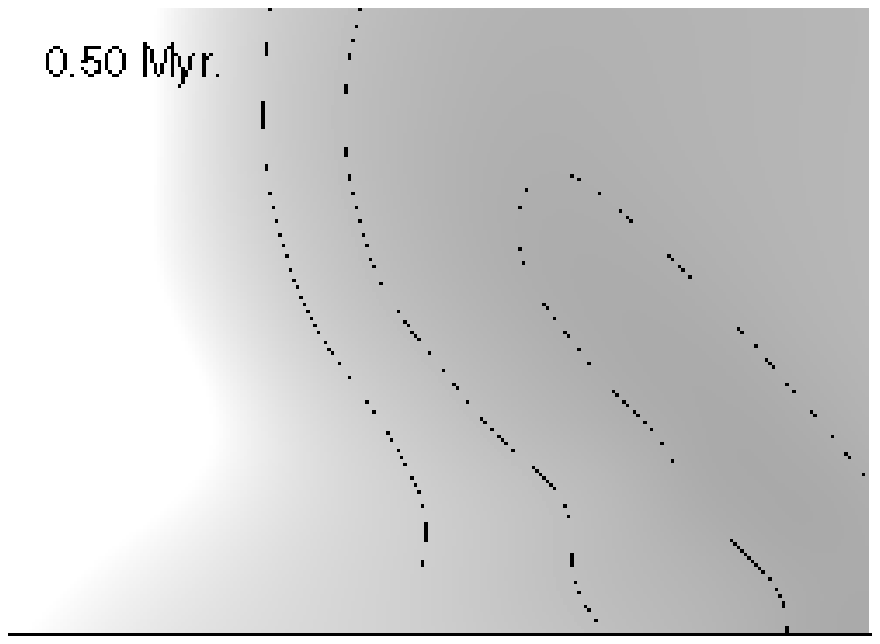}{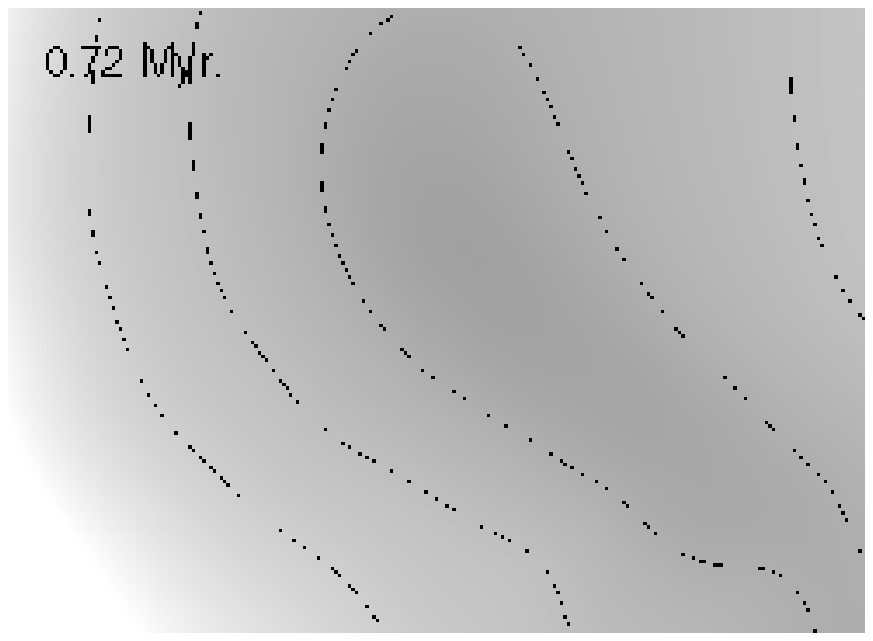}\\
    \vspace{6pt}
    \plottwo{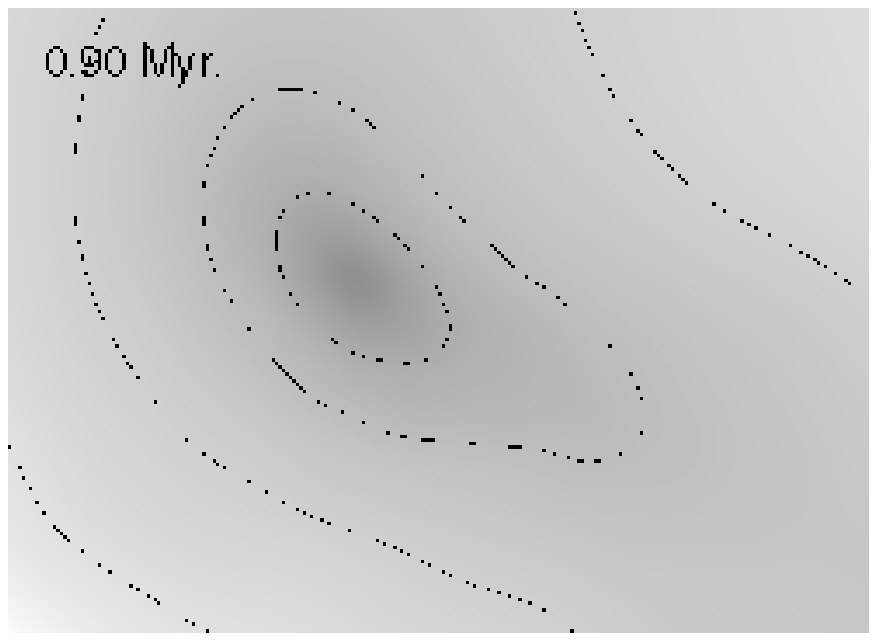}{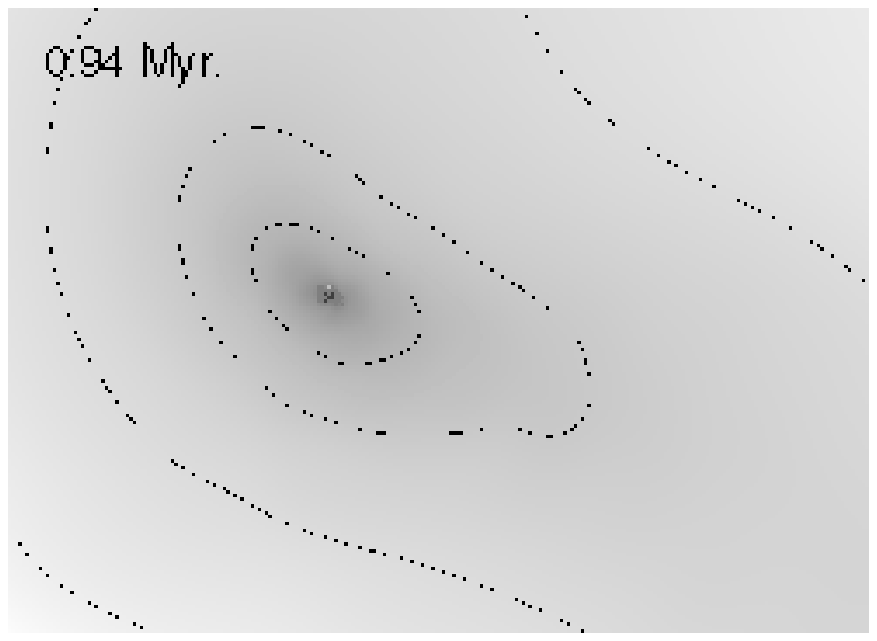}\\
    \vspace{6pt}
    \plottwo{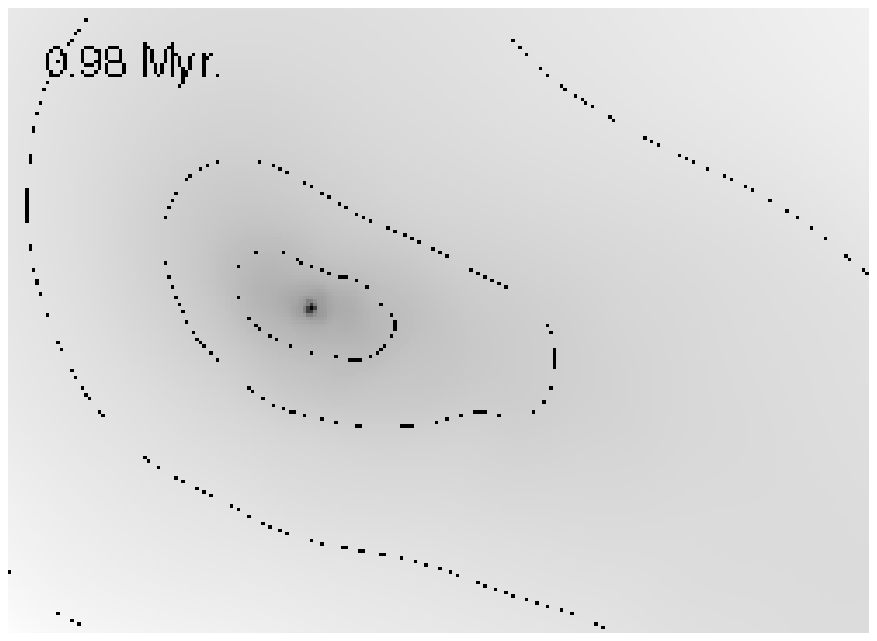}{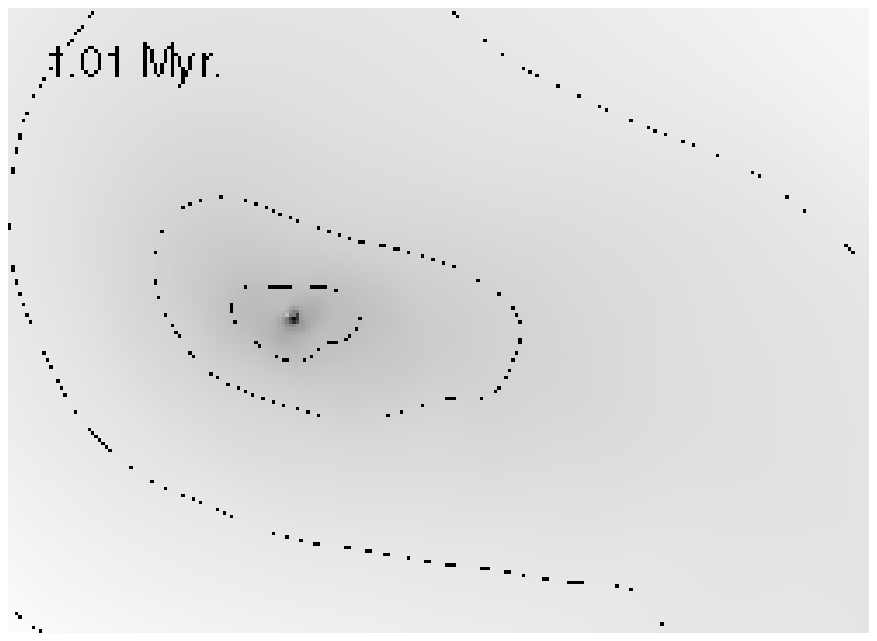}\\
  \caption{$b$=0.7 collision evolution (upper left region shown, lower right is very similar). The collision perturbs and compresses the leading edge of each clump, causing collapse. [$x$=(-0.60,0.00)pc, $y$=(0.00,0.45)pc, contours at $\logn$ = (21.6, 21.9, 22.3, 22.6), min=-$\infty$, max=25.9.]}
  \label{fig:cc-b0.7evolve}
  \end{minipage}
\end{figure*}

The collision produced a diagonal density enhancement (Fig. \ref{fig:cc-b0.7evolve}), with the enhancement strongest at the ends of the diagonal. These regions subsequently collapsed to produce two well separated condensations, and whilst the condensations were forming the leading edge of each clump expanded into the rarefied region behind the trailing edge of the other clump. When the simulation was terminated at 1 Myr two condensations of mass 7$\smass$ and radius $\sim$660 AU were present, at a separation of 0.94 pc. This collision is in a relatively early stage of evolution, as the condensations form at later times than those in low impact parameter collisions.

\subsection{Dependence on the impact parameter.}
\label{subsec:dependence}

\begin{figure*}
  \begin{minipage}{150mm}
  \plotxsize{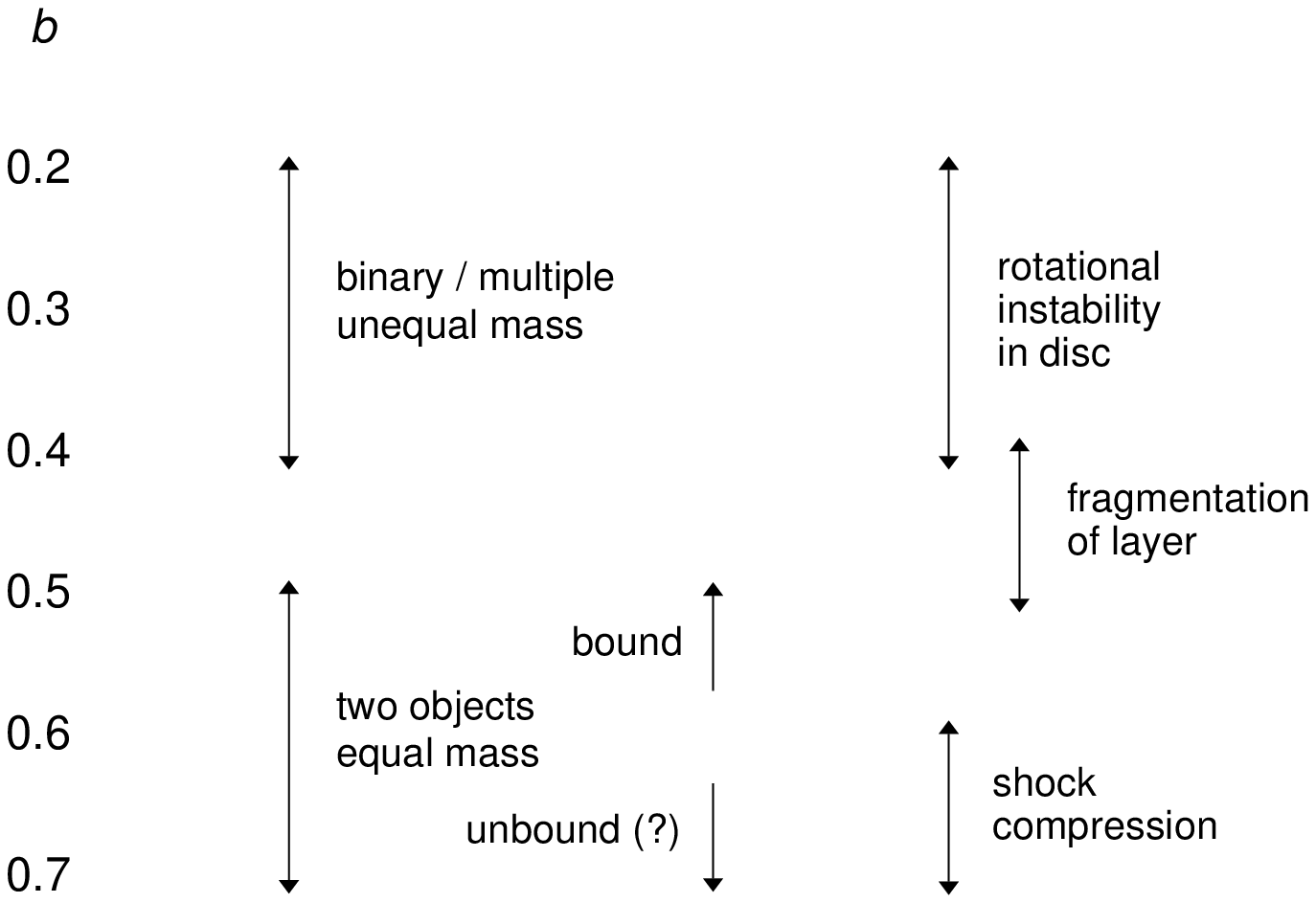}{100mm}
  \caption{Summary of the results and the disc formation mechanisms which occur in the clump-clump collision simulations, as a function of the impact parameter $b$. See \S\ref{subsec:dependence} for a fuller discussion.}
  \label{fig:cc-bscansum}
  \end{minipage}
\end{figure*}

The main points of this discussion are summarised in Fig. \ref{fig:cc-bscansum}.

Low $b$ collisions ($b$=0.2, 0.3) produce multiple objects by a process of accretion induced rotational fragmentation acting on an initially formed central condensation (see also Whitworth et al \shortcite{whi-arf}). Unequal mass components result from this process, at least up to the end of the simulations presented here (1 Myr after the collision). Decreasing the impact parameter produces a higher mass primary ($M_{1}$=60$\smass$ if $b$=0.2, and 40$\smass$ if $b$=0.3) since the reduced clump offset results in incoming material having less angular momentum. This slows the spin-up and allows more mass to accrete onto the primary. 

A collision of $b$=0.4 includes mechanisms present both in the low $b$ and $b$=0.5 collisions, with the accretion flow onto the central disc eventually fragmenting.

Higher $b$ collisions ($b$=0.5, 0.6, 0.7) produce two equal mass protostellar objects in separate regions. The formation times and separations increase as $b$ is increased. The systems are not well evolved by the end of the simulations, and the two objects in each collision have not interacted. However, the final objects in the $b$=0.5 collision are bound, those in the $b$=0.6 case are marginally bound, and those in the $b$=0.7 case are unbound. The objects in the two high impact parameter collisions will probably go on to accrete significant amounts of mass, and thus they may subsequently become bound. The ratio of the two condensation's mutual kinetic energy to gravitational potential energy is $\sim 4.6$ at the end of the simulation.

The $b$=0.5 collision produces its protostellar discs by a process of shock-induced gravitational fragmentation, followed by capture and merger between fragments (see also Turner et al \shortcite{tur}).

The high impact parameter collisions ($b$=0.6, 0.7) produce objects at the leading edges of the clumps due to the compression of the impact triggering collapse. The collapse occurs because pressure support is reduced in the compressed regions as the gas cools.

If $\bsw$ is defined as the impact parameter below which rotational instabilities are dominant, then $\bsw \sim$0.45. Unequal mass systems result if $b<\bsw$, and equal mass systems (i.e. $M_{1} \sim M_{2}$) if $b>\bsw$. It is expected that the binary systems produced when $b<\bsw$ will have smaller separations than those when $b>\bsw$, since although the binary components of the latter are bound at formation they initially have much larger separations.

\section{Varying numerical noise.}
\label{sec:orient}

\begin{figure*}
  \begin{minipage}{150mm}
    \plotps    {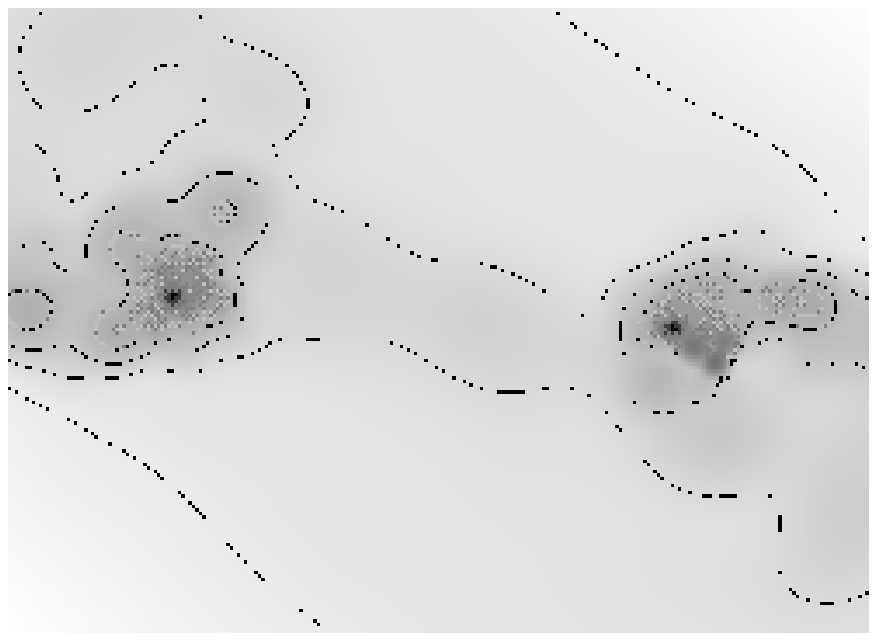}
    \plotysize {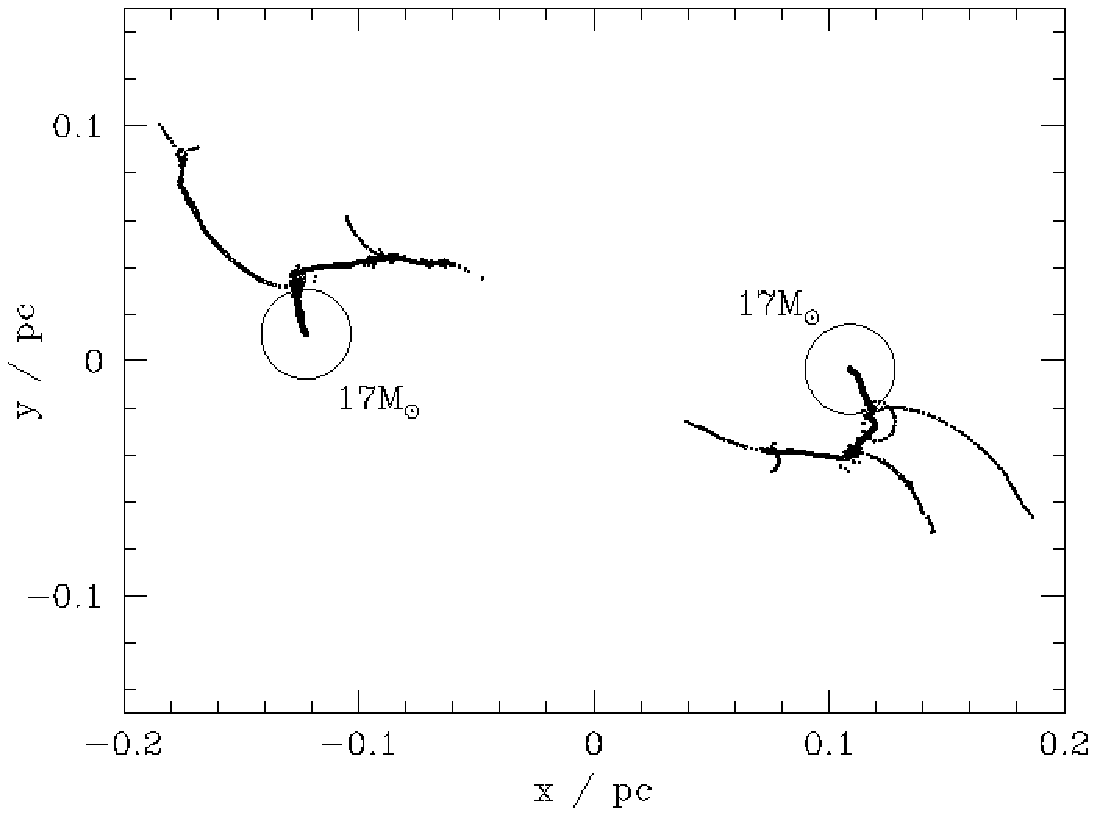}{58mm} \\
    \vspace{6pt}
    \plotps    {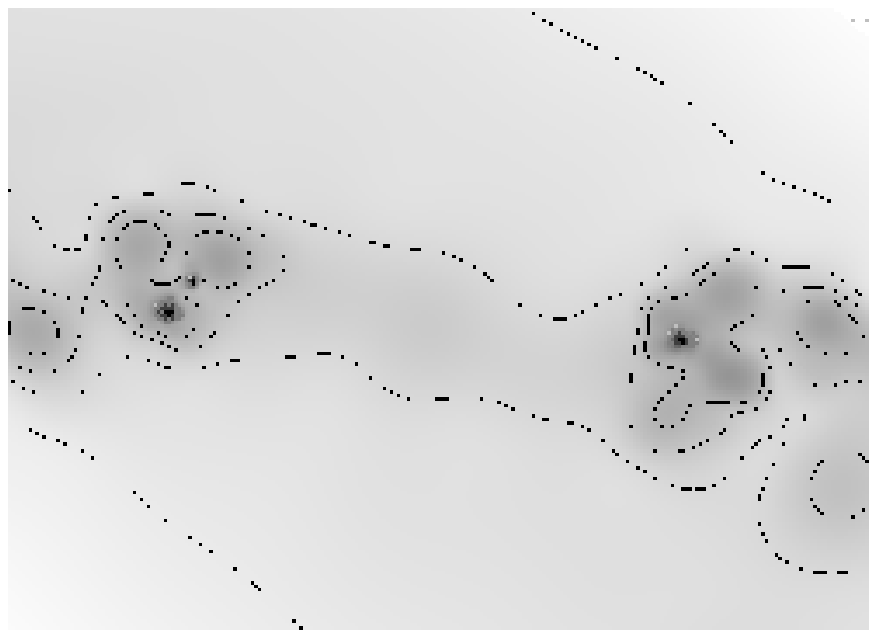}
    \plotysize {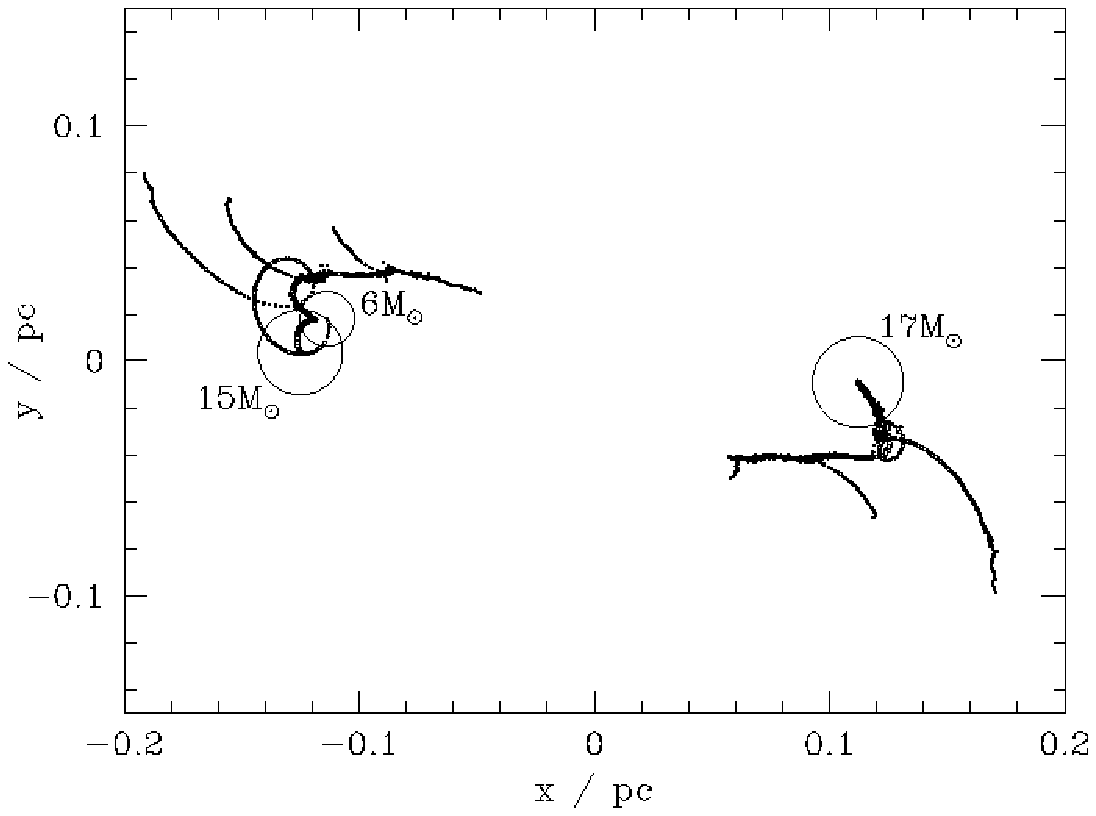}{58mm} \\
    \vspace{6pt}
    \plotps    {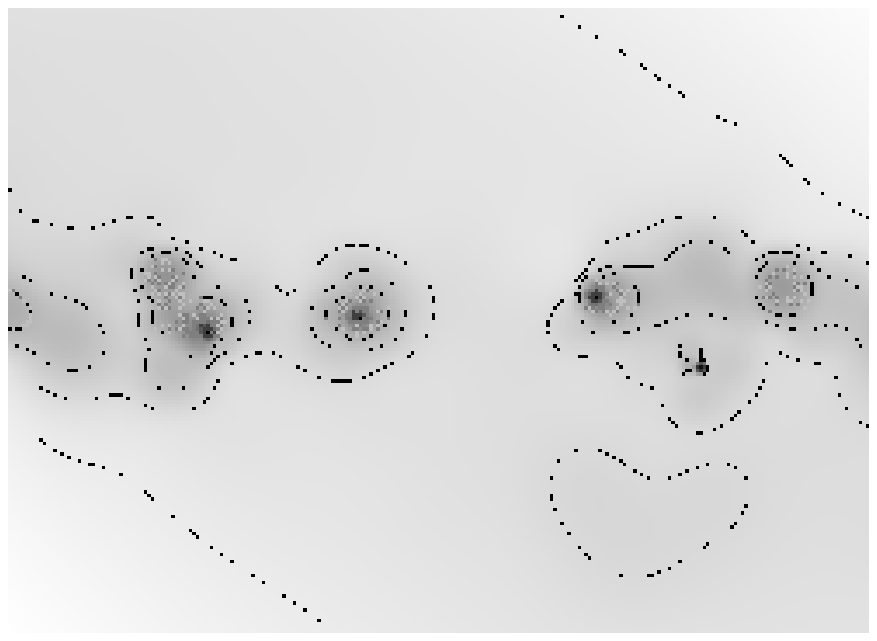}
    \plotysize {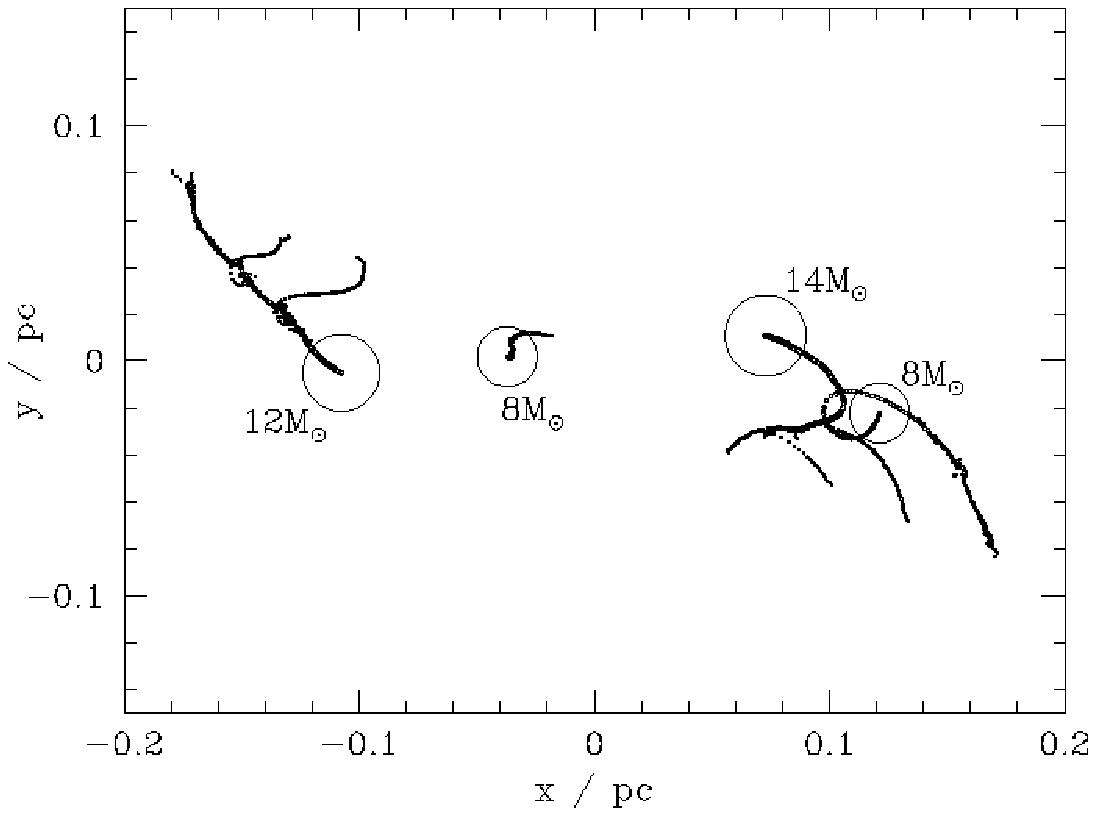}{58mm} \\
  \caption{Re-oriented, $b$=0.5 collisions. Final states : original collision (top), clumps re-oriented by 120$\deg$ about the $z$-axis (middle), and by 240$\deg$ about the $z$-axis (bottom). Left : column density. [$x$=(-0.20,0.20)pc, $y$=(-0.15,0.15)pc, contours at $\logn$ = (21.9, 22.3, 22.6, 22.9), min=21.6, max=26.4.] Right : paths taken by condensations throughout the simulation, and final objects shown as circles with area proportional to mass.}
  \label{fig:cc-orifinal}
  \end{minipage}
\end{figure*}

\begin{table*}
\begin{minipage}{150mm}
\caption{Summary of the re-oriented clump collisions (see \S\ref{sec:orient}).}
\label{table:orient}
\begin{tabular}{c|cccccccc}
$b$ & $t_{*}$ & $r_{*}$ & $n_{*}$ & $M_{*}$ & $j_{*}$ & $\Delta L_{z}$\\
\footnotesize (/dia) & \footnotesize (Myr) & \footnotesize (AU) & \footnotesize (10$^{11}$cm$^{-3}$) & \footnotesize ($\smass$) & \footnotesize (10$^{24}$cm$^{2}$s$^{-1}$) & \footnotesize (\%) \\ \hline
& & & & & & & & \\
0.5 & 0.78 & 890,920 & 1.2,1.5 & 17,17 & 1.3,1.3 & -1 \\
& \multicolumn{7}{l}{\footnotesize Rotated by 0$^{\circ}$ (standard collision). Two objects, each formed by multiple fragment mergers. \normalsize}\\
& & & & & & & & \\
0.5 & 0.76 & 890,780 & 1.1,1.5 & 15,17 & 1.2,1.3 & -2 \\
& \multicolumn{7}{l}{\footnotesize Rotated by 120$^{\circ}$. Three objects (6-17$\smass$), two dominant formed by multiple fragment mergers. \normalsize}\\
& & & & & & & & \\
0.5 & 0.76 & 790,750 & 0.9,1.3 & 12,14 & 1.1,1.0 & -2 \\
& \multicolumn{7}{l}{\footnotesize Rotated by 240$^{\circ}$. Four objects (8-14$\smass$), formed by multiple fragment mergers. \normalsize}\\
& & & & & & & & \\
\end{tabular}
$r_{*}$, $M_{*}$ and $j_{*}$ are determined using particles above a density threshold of $n=1.7\times10^{8}$ cm$^{-3}$. $t_{*}$ is the earliest time 2$\smass$ of material forms a compact condensation. $n_{*}$ is the maximum density of a protostar at the end of the simulation.
\end{minipage}
\end{table*}

Two additional collisions were performed which were identical to the standard $b$=0.5 collision, but with each clump re-oriented before the collision. This allowed the dependence of the results on particle noise to be investigated (the noise is a result of the discrete nature of the particle distribution). Since the clumps used have been relaxed, there are no grid-aligned directions, or any other preferential axes. Both clumps were rotated about an axis passing through their centres in the $z$-direction by $120^{\circ}$ for one of the collisions, and by $240^{\circ}$ for the other. The value $b$=0.5 was chosen because it was thought to be the impact parameter most sensitive to noise.

The final states of the original $b$-scan collision plus the two new orientations are shown in Fig. \ref{fig:cc-orifinal}, and are summarised in Table \ref{table:orient}. Multiple fragmentation and subsequent mergers occurred in each collision, usually until two pairs of objects remained at each end of the shock. Merger between each of the objects in the pairs was a possibility. At the right-hand side of the $120^{\circ}$ case a 8$\smass$ secondary merged with a 12$\smass$ primary after completing about one orbit. Material was ejected during the merger process, and a single 17$\smass$ object was produced. At the left-hand side merger did not occur, with a 6$\smass$ object thrown into an elliptical orbit around a 15$\smass$ primary. More than one orbit was completed. Merger also did not occur at the right-hand side of the $240^{\circ}$ case, where a 7$\smass$ secondary encountered a 14$\smass$ primary. Instead, momentum was exchanged and the primary was ejected towards the collision centre. Merger of small fragments occured at the left-hand side to produce a 12$\smass$ object, and also just left of the centre producing another 8$\smass$ object.

The shock size, shock shape, time of fragmentation, the total mass of discs produced, and the regions in which these discs appeared were broadly similar across all the collisions. The differences concerned where fragmentation occurred, and whether final merger events took place. Fragmentation was usually localised within two regions in each collision. These areas were small enough to allow the objects produced within each region to interact, and to merge until the last pair of objects remained. By this stage the objects had evolved into discs, and the probability of merger was found to be $\sim$50\%. In one of the collisions however, fragmentation also occurred in a more central region, and a single object was produced. The central disc was at a sufficient distance from the other fragmentation regions to preclude interaction up to 1Myr after the time of collision.

\section{Varying numerical resolution.}
\label{sec:hires}

To investigate the effect of resolution on the outcome of a simulation, the standard $b$=0.5 collision was repeated with the number of particles in each clump increased by a factor of four. This increased the linear spatial resolution by a factor of  $4^{\frac{1}{3}}$, or $\sim$1.6. Accordingly, the gravity smoothing length $\epsilon$ was dropped by the same factor, from 0.002 pc to 0.00127 pc. A similar change was not required for the hydrodynamic smoothing length $h$, since the size is controlled by the requirement that a reasonable number of neighbours be found within each kernel (this automatically reduces $h$ to a suitable value). A total of 95411 particles were used (7999 per clump, 52031 representing the interclump, and 27382 forming the boundary).

\begin{table*}
\begin{minipage}{150mm}
\caption{Summary of the high-resolution collision (see \S\ref{sec:hires}).}
\label{table:hires}
\begin{tabular}{c|cccccccc}
$b$ & $t_{*}$ & $r_{*}$ & $n_{*}$ & $M_{*}$ & $j_{*}$ & $\Delta L_{z}$\\
\footnotesize (/dia) & \footnotesize (Myr) & \footnotesize (AU) & \footnotesize (10$^{11}$cm$^{-3}$) & \footnotesize ($\smass$) & \footnotesize (10$^{24}$cm$^{2}$s$^{-1}$) & \footnotesize (\%) \\ \hline
& & & & & & & & \\
0.5 & 0.77 & 560,580 & 3.4,4.7 & 11,12 & 0.7,0.8 & -2 \\
& \multicolumn{7}{l}{\footnotesize Six objects (1-12$\smass$). Two binaries (9,12 \& 11,12$\smass$), components form by mult. frag. mergers.\normalsize}\\
& & & & & & & & \\
\end{tabular}
$r_{*}$, $M_{*}$ and $j_{*}$ are determined using particles above a density threshold of $n=1.7\times10^{8}$ cm$^{-3}$. $t_{*}$ is the earliest time 2$\smass$ of material forms a compact condensation. $n_{*}$ is the maximum density of a protostar at the end of the simulation.
\end{minipage}
\end{table*}

The collision evolution followed a similar pattern to the equivalent standard-resolution collision, but the shock produced was thinner, and the fragmentation was better resolved. The result is summarised in Table \ref{table:hires}.

\begin{figure*}
  \begin{minipage}{150mm}
    \plotps    {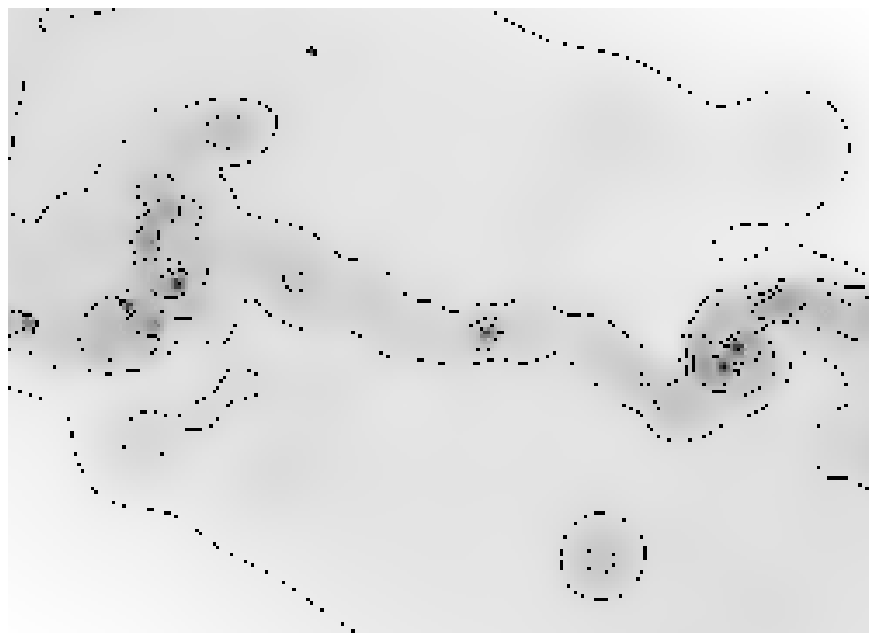}
    \plotysize {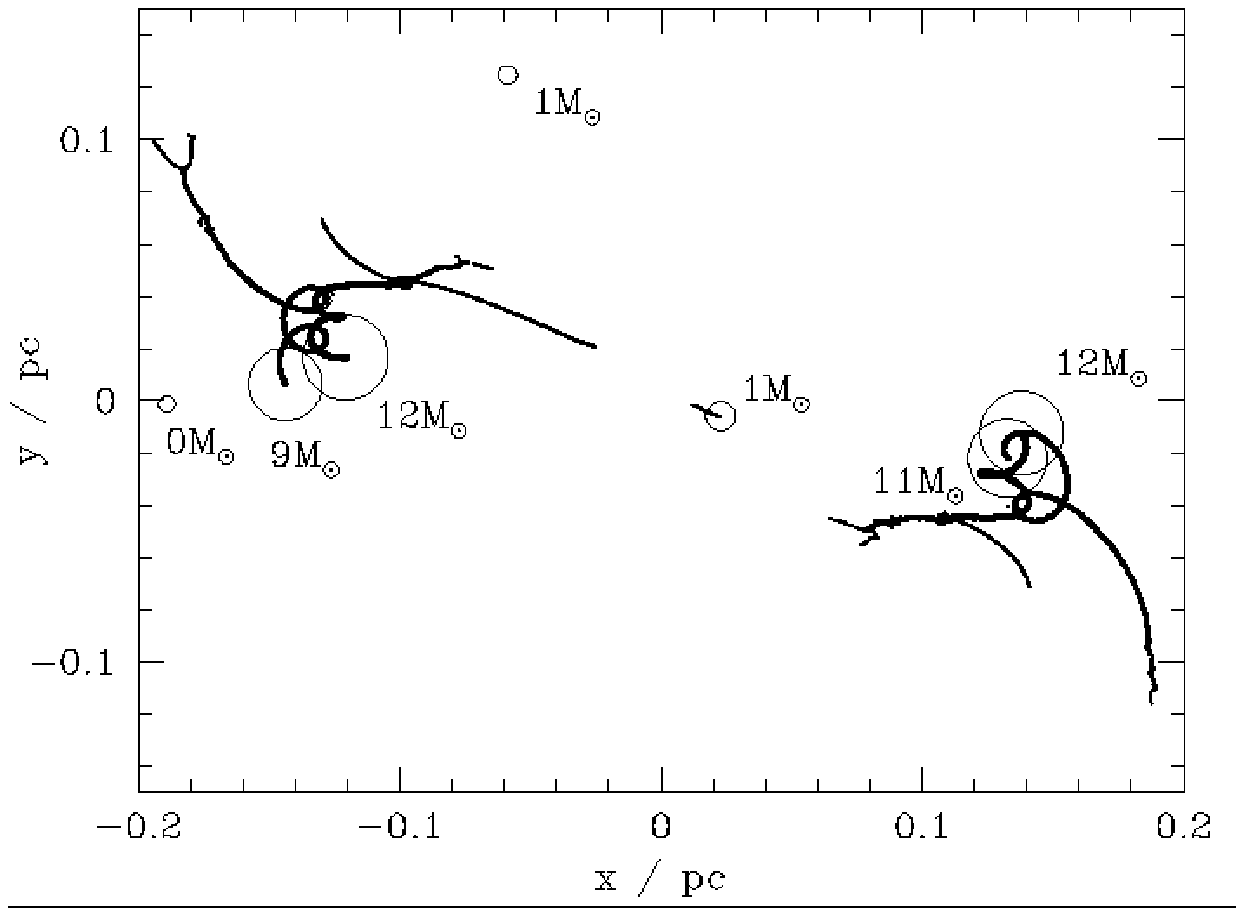}{58mm} \\
  \caption{High-resolution, $b$=0.5 collision. Final state. Left : column density. [$x$=(-0.20,0.20)pc, $y$=(-0.15,0.15)pc, contours at $\logn$ = (21.9, 22.3, 22.6, 22.9), min=21.6, max=26.7.] Right : paths taken by condensations of 1$\smass$ and above throughout the simulation, and final objects shown as circles with area proportional to mass. Although the high-resolution simulation captures more detail of early low-mass objects than the standard-resolution simulation, the paths taken by subsequent objects (shown here) are similar (though final mergers do not occur and binaries are the end result).}
  \label{fig:cc-hirfinal}
  \end{minipage}
\end{figure*}

Fragmentation initially occurred near to the collision centre within the shock bounded region. As the collision progressed, the density in regions of the shock further from the centre increased, and fragmentation spread out along the shock. By 0.92 Myr fragment merger had produced an approximately symmetric configuration with two pairs of discs on either side of a small 1$\smass$ central condensation. The inner discs formed from mergers close to the collision centre, and had masses of 11$\smass$ each. The outer discs formed from mergers in the outer regions of the shock, and were slightly smaller (each had a mass of 8$\smass$). This configuration is similar to that found in the various standard-resolution $b$=0.5 collisions before the final encounters at either side of the shock take place. Here, the two small discs encountered the larger discs at 0.95 Myr and merger did not occur in either encounter. Instead, matter was ejected via the production and subsequent detachment of tidally induced arms (which also removed angular momentum and increased the binding energy of each pair of discs), and binary systems resulted. The final state of the system is shown in Fig. \ref{fig:cc-hirfinal} (left).

When compared to the standard-resolution simulations, some differences of detail are apparent : the binary components are of slightly smaller mass (9-12$\smass$ instead of 12-17$\smass$), and more significantly the objects have higher peak densities, less specific spin angular momentum, and are smaller (at $<$600 AU instead of 800-900 AU). Thus the objects produced in the high-resolution simulation are more tightly bound (a consequence of the reduction in $\epsilon$ and $h$). This reduction in size and increase in density is apparent throughout the simulation and causes the discs in the final encounters to resist merger, and to favour capture. Since the high resolution simulation is more realistic, it is expected that the production of binaries would be favoured in nature.

A plot of fragment paths reveals many more objects in the high-resolution simulation. However, since a density threshold is applied, this means that {\it denser} fragments are more common. If in addition a 1$\smass$ threshold is applied (Fig. \ref{fig:cc-hirfinal}, right) then the plot closely resembles the standard-resolution plots (Fig. \ref{fig:cc-orifinal}, these are left almost unaltered by applying the mass threshold). Thus increasing the resolution allows better simulation of early small and dense fragments, but these subsequently merge and the larger scale behaviour is little changed.

\begin{figure*}
  \begin{minipage}{150mm}
    \plotthree{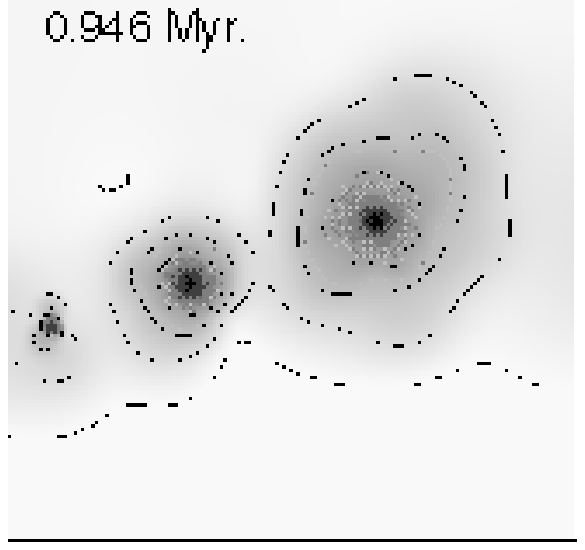}{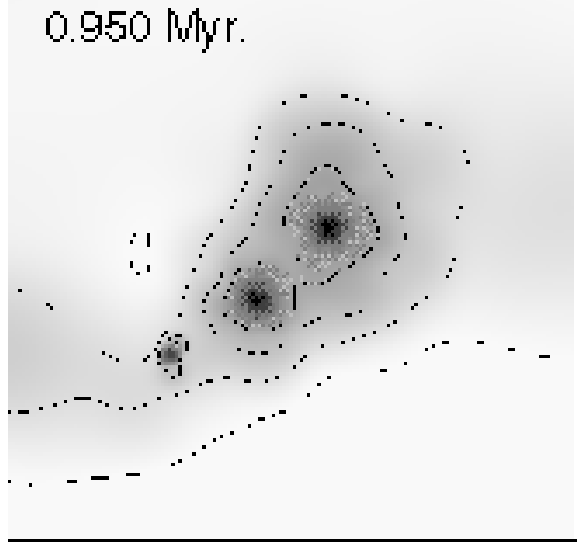}{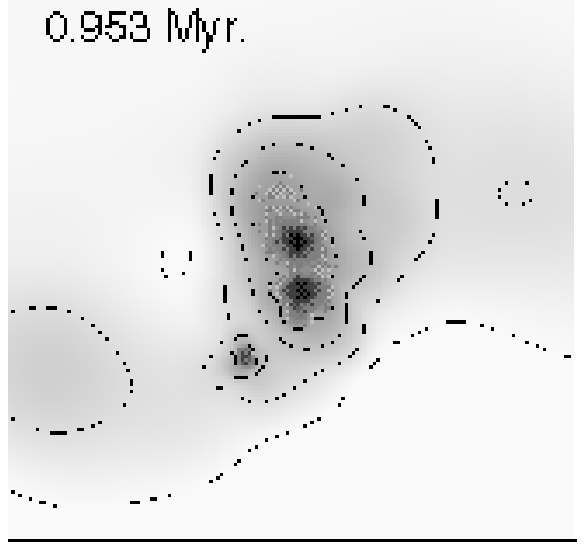} \\ 
    \vspace{6pt}
    \plotthree{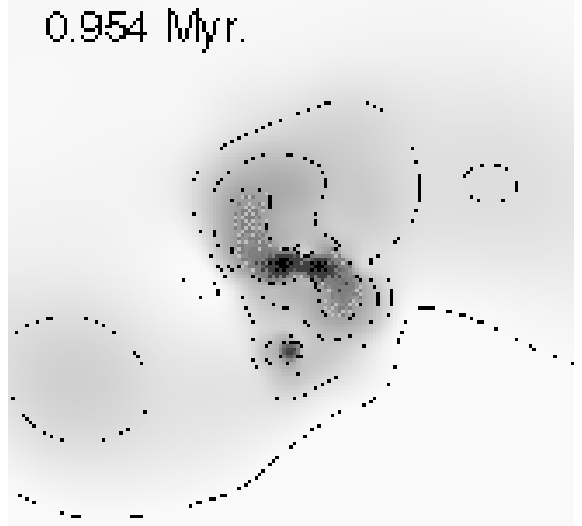}{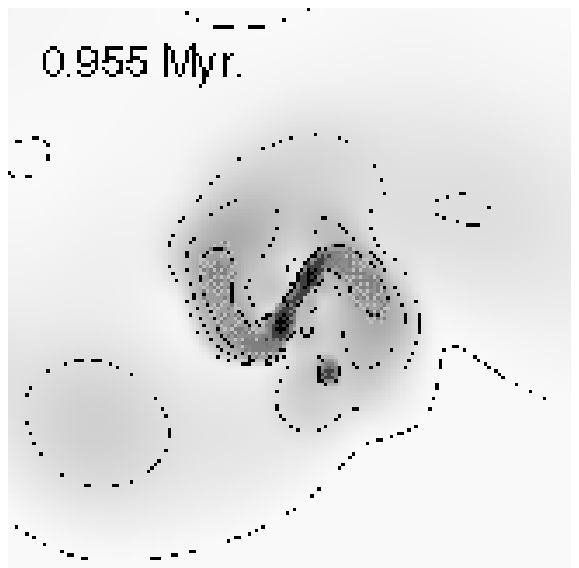}{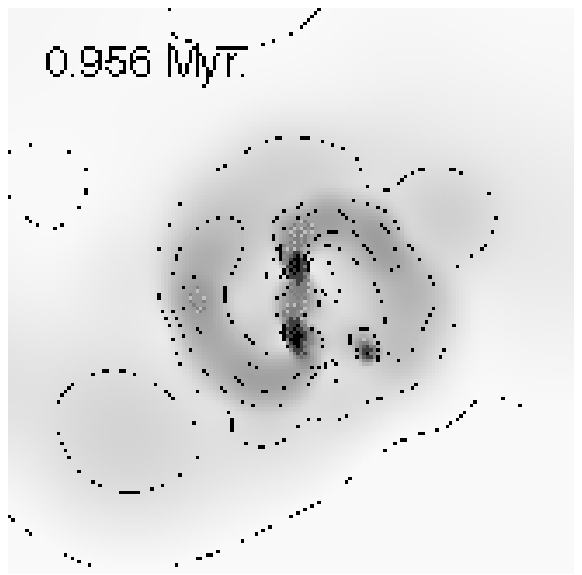} \\ 
    \vspace{6pt}
    \plotthree{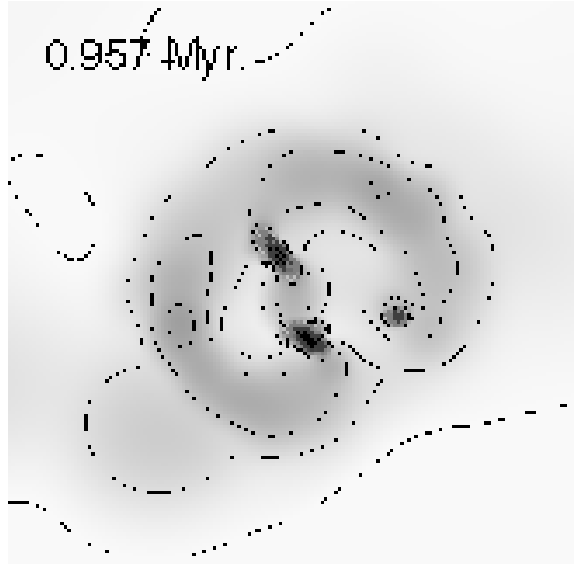}{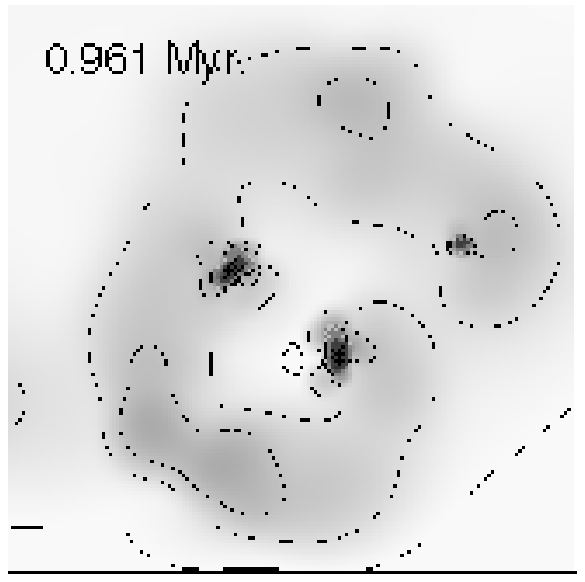}{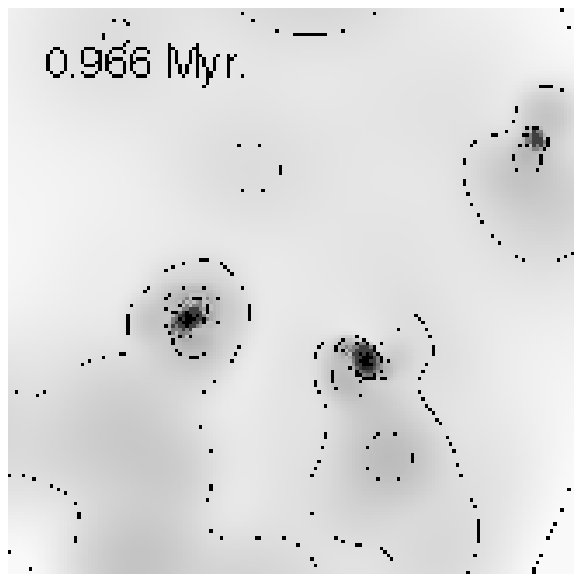} \\ 
  \caption{High-resolution, $b$=0.5 collision. Encounter between two discs (8 and 11 $\smass$), and a 1$\smass$ fragment. The fragment is ejected from the three-body system to leave a binary system. [$x$=(-0.165,-0.097)pc, $y$=(0.006,0.074)pc, contours at $\logn$ = (22.2, 22.7, 23.2, 23.7).]}
  \label{fig:cc-hrbinary}
  \end{minipage}
\end{figure*}

Another small scale difference was the ability to model late forming low-mass fragments appearing at the outer edges of the shock, far from the collision centre. These fragments were few in number, and their subsequent encounters with the binaries usually resulted in disruption or merger with one of the binary discs. However, one of the low-mass fragments encountered the two left-hand discs just as they were at periastron during their first encounter. A three-body event occurred and the 1$\smass$ fragment was ejected. Fig. \ref{fig:cc-hrbinary} shows a time sequence of this encounter showing the ejection of this small object, and the tidal arms excited in the main discs. This is a process by which low mass objects may survive, potentially creating lower mass stars.

\section{Conclusions.}
\label{sec:conclude}

The parameter space investigated was restricted to clumps of {\it equal} mass. We believe that relaxing this requirement to consider collisions between clumps of {\it similar} mass (which probably encompasses the majority of collisions occurring in giant molecular clouds), will produce results not greatly different to those presented here. Larson's relations \cite{lar} suggest that clump radius and density vary slowly with mass ($r \propto M^{1/2}$, $\rho \propto M^{-1/2}$), and so the geometry of such collisions will be largely unaltered. However, the result of any collisions that occur between clumps of widely differing mass may well be significantly different, since both the geometry and the extent of the shocked region will be substantially changed. 

The internal properties of the discs produced here, such as density and velocity profiles, cannot be determined reliably. This is due both to the excessive shear viscosity introduced by the SPH artificial viscosity formulation, and the comparatively large size of $\epsilon$ required to prevent infinite density collapse. A small $\epsilon$ also results in a prohibitively small timestep).

Unfortunately, it is also not possible to follow the subsequent evolution of the protostellar objects formed in these simulations. This evolution seems likely to involve a competition between the internal evolution of the individual protostellar discs (which converts a massive protostellar disc into a central star plus an emaciated circumstellar disc) and impulsive interactions between the discs (which may cause them to fragment and spawn additional protostellar discs in closer binary systems). In order to explore this possibility a new SPH viscosity formulation has been developed in which the bulk and shear components can be regulated independently \cite{wat:vis}. Simulations which concentrate on examining solely the interaction between a protostellar disc and either a naked star or another protostellar disc, have been performed using this method \cite{wat:dis}. Such interactions are found to be effective at spawning new protostars and closer binary systems.

The larger scale processes which deliver the material that produces the discs in the simulations presented here are believed to be well resolved. Thus, it seems that the mechanisms which produce the discs are viable, and processes such as fragmentation and accretion-induced rotational instabilities are well modelled. The relatively large value of $\epsilon$ (compared to the typical disc radius) produces discs which are more susceptible to disruption than they would be otherwise (since they are less well bound). The presence of many surviving discs, in spite of the high $\epsilon$, indicates that disc formation is a likely consequence of a clump-clump collisions in nature. In addition, quantites such as the masses, positions and orientations of the discs are expected to be unaffected by numerical factors, and so are believed to be reliable.

Low impact parameter collisions ($b$=0.2, 0.3) produced binary or multiple systems by the rotational instability of an initially formed central disc. In these systems the components tended to have unequal masses. Mid-$b$ collisions produced two bound objects of similar mass which formed via repeated mergers of fragments which appeared in the shock-compressed layer produced at the collision interface. High-$b$ collisions ($b$=0.6, 0.7) produced discs, also of similar mass, at the leading edges of each clump - these were unbound in the $b$=0.7 case but were still accreting significant amounts of matter at the simulation end (1 Myr after collision). If the simulations could be followed further, a larger fraction of the mass would end up in protostellar discs. However, this is because we have not included any feedback processes, of the sort which might cut off accretion onto a protostellar disc.

The numerical noise introduced by discretising the gas into a system of particles did not appear significantly to affect the results. Collisions between re-oriented clumps produced results broadly similar to the original collision simulation.

An increase in resolution also left the results of a simulation largely unchanged - early small scale fragmentation was better resolved but subsequent evolution was left largely unaltered, though the final discs were slightly smaller and of slightly higher density.

Although not presented here, collisions between spinning clumps were also investigated. The effect of a spin rate similar to that observed for such clumps was found to be negligible - the kinetic energy and angular momentum associated with the spins of the individual clumps are much smaller than the translational kinetic energy and the orbital angular momentum associated with the collision.

It is concluded that clump-clump collisions provide a realistic mechanism by which protostars may be formed. In fact, binary or multiple systems are often the result; this is in good agreement with observations of star-forming regions, which similarly find a preponderance of multiples and binaries. Collisions with a wide range of impact parameters were found to produce two or more protostellar objects.

\end{document}